\documentclass[aps, prd, 10pt, twocolumn, superscriptaddress, noshowpacs, preprintnumbers, longbibliography, groupedaddress, bibnotes,hyperref]{revtex4-1}

\usepackage{amsmath}
\usepackage{amsfonts}
\usepackage{amssymb}
\usepackage{bbold}
\usepackage{epsfig}
\usepackage{graphicx}
\usepackage{bm}
\usepackage{array}
\usepackage{hyperref}
\usepackage{listings}
\usepackage{color}
\usepackage{float}
\usepackage[normalem]{ulem}

\newcommand{\cemit}{\mathcal{C}_{\textrm{emission}}}
\newcommand{\cabsorb}{\mathcal{C}_{\textrm{absorb}}}
\newcommand{\cdirch}{\mathcal{C}_{\textrm{dir-ch}}}

\usepackage{tikz,xcolor,hyperref}

\definecolor{lime}{HTML}{A6CE39}
\DeclareRobustCommand{\orcidicon}{\hspace{-1mm}
	\begin{tikzpicture}
	\draw[lime, fill=lime] (0,0) 
	circle [radius=0.16] 
	node[white] {{\fontfamily{qag}\selectfont \tiny \,ID}};
	\draw[white, fill=white] (-0.0525,0.095) 
	circle [radius=0.007];
	\end{tikzpicture}
	\hspace{-3mm}
}

\foreach \x in {A, ..., Z}{\expandafter\xdef\csname orcid\x\endcsname{\noexpand\href{https://orcid.org/\csname orcidauthor\x\endcsname}
			{\noexpand\orcidicon}}
}

\begin{document}

\title{Neutrino Flavor Conversion, Advection, and Collisions: Towards the Full Solution}
\author{Shashank Shalgar\orcidA{}}
\affiliation{Niels Bohr International Academy \& DARK, Niels Bohr Institute,\\University of Copenhagen, Blegdamsvej 17, 2100 Copenhagen, Denmark}

\author{Irene Tamborra\orcidB{}}
\affiliation{Niels Bohr International Academy \& DARK, Niels Bohr Institute,\\University of Copenhagen, Blegdamsvej 17, 2100 Copenhagen, Denmark}

\date{\today}

\begin{abstract}
At high densities in compact astrophysical sources, the coherent forward scattering of neutrinos onto each other is responsible for making the flavor evolution non-linear. Under the assumption of spherical symmetry, we present the first simulations tracking flavor transformation in the presence of neutrino-neutrino forward scattering, neutral and charged current collisions with the matter background, as well as neutrino advection. We find that, although flavor equipartition could be one of the solutions, it is not a generic outcome, as often postulated in the literature. Intriguingly, the strong interplay between flavor conversion, collisions, and advection leads to a spread of flavor conversion across the neutrino angular distributions and neighboring spatial regions. Our simulations show that slow and fast flavor transformation can occur simultaneously. In the light of this, looking for crossings in the electron neutrino lepton number as a diagnostic tool of the occurrence of flavor transformation in the high-density regime is a limiting method. 

\end{abstract}

\maketitle

\section{Introduction}
\label{sec:intro}
In neutrino-dense sources, such as core-collapse supernovae, 
the large number of neutrinos drives the source physics,  despite the weakness of their interaction~\cite{Burrows:2020qrp,Janka:2016fox, 
Bethe:1985sox, 1966ApJ...143..626C,1985nuas.conf..422W}.  
Tracking the neutrino flavor evolution in the source core  is a complex task because, in addition to resonant conversion of neutrinos in matter~\cite{1985YaFiz..42.1441M, 1978PhRvD..17.2369W}, the  coherent forward scattering of neutrinos on other neutrinos makes the flavor evolution  non-linear~\cite{Pantaleone:1992eq, Sigl:1992fn, Duan:2010bg,Mirizzi:2015eza,Tamborra:2020cul}.  
Neutrino self-interaction is a   peculiar phenomenon: neutrinos with different momenta undergo  flavor evolution with identical characteristic frequency~\cite{Duan:2007bt,Duan:2006jv, Duan:2006an,Duan:2005cp,Fogli:2007bk,Fogli:2008pt,Raffelt:2007cb,Hannestad:2006nj}.

In the context of core-collapse supernovae, collective neutrino oscillation was originally conceptualized within the ``neutrino-bulb'' model~\cite{Duan:2006an}. 
The latter was based on the assumption of spherical symmetry and instantaneous decoupling of all neutrino flavors at a single radius. Within this framework, neutrino self-interaction leads to a swap in the energy distributions of the electron and non-electron flavors, the spectral split~\cite{Duan:2007bt, Duan:2006an,Fogli:2007bk,Fogli:2008pt,Raffelt:2007cb,Dasgupta:2009mg,Fogli:2009rd,Dasgupta:2010cd,Friedland:2010sc}.  However, soon it was realized that the non-linear nature of neutrino collective effects leads to spontaneous breaking of symmetries~\cite{Raffelt:2013rqa, Duan:2014gfa}.

Neutrinos of different flavors interact with matter differently. As a consequence, 
a crossing between the angular distributions of electron neutrinos and antineutrinos, the electron lepton number (ELN) crossing, may occur~\cite{Shalgar:2019kzy,Nagakura:2021hyb}. Because of this feature, neutrinos  experience a flavor instability, also in the limit of vanishing vacuum frequency~\cite{Sawyer:2005jk,Sawyer:2008zs,Sawyer:2015dsa, Chakraborty:2016yeg, Chakraborty:2016lct, Tamborra:2020cul,Izaguirre:2016gsx,Yi:2019hrp,Martin:2019gxb}. The flavor transformations resulting from such a flavor instability can occur at arbitrarily large neutrino number densities~\cite{Sawyer:2005jk, Sawyer:2015dsa, Tamborra:2020cul, Izaguirre:2016gsx, Shalgar:2020xns, Shalgar:2019qwg, Shalgar:2021wlj, Johns:2019izj, Chakraborty:2019wxe, Abbar:2018shq, Dasgupta:2016dbv}. The corresponding characteristic frequency associated with flavor transformation can be very large, lending the phenomenon the name of ``fast flavor conversion'' to distinguish it from the ordinary neutrino self-interaction~\cite{Chakraborty:2016yeg,Tamborra:2020cul}. The latter is also named ``slow''  collective oscillation since it is governed by a combination of the neutrino-neutrino self-interaction potential and the vacuum frequencies~\cite{Duan:2010bg,Mirizzi:2015eza}.

Despite being driven by the angular distributions of neutrinos, fast flavor conversion is further affected by the vacuum term and by the presence of all three neutrino flavors~\cite{Shalgar:2020xns,Chakraborty:2019wxe,Capozzi:2020kge,Shalgar:2021wlj,Capozzi:2022dtr}, as well as by symmetry breaking effects~\cite{Shalgar:2021oko}. Moreover,   
collisions can enhance or suppress  fast flavor transition according to the neutrino angular distributions~\cite{Shalgar:2020wcx,Johns:2021qby,Hansen:2022xza,Johns:2022bmu,Martin:2021xyl,Shalgar:2022rjj}. 
Unlike in the case of the neutrino-bulb model, which ignores the temporal dependence of the neutrino field due to advection, the motion of neutrinos can alter their momentum distribution; this can, in turn, affect the flavor evolution~\cite{Shalgar:2019qwg, Richers:2021xtf, Nagakura:2022kic,Shalgar:2022rjj}. 

Favorable conditions for the occurrence of fast flavor instabilities have been found in core-collapse supernovae as well as in compact binary merger remnants~\cite{Xiong:2020ntn, Wu:2017drk, Just:2022flt,George:2020veu,Li:2021vqj, Tamborra:2017ubu, Shalgar:2019kzy, Abbar:2018shq, Azari:2019jvr, DelfanAzari:2019tez,DelfanAzari:2019epo,Morinaga:2019wsv, Glas:2019ijo, Abbar:2019zoq, Nagakura:2019sig,Abbar:2020qpi,Capozzi:2020syn,Nagakura:2021hyb, Harada:2021ata}. These developments have triggered intense research work aiming to assess the feedback of flavor transformation on the source physics. However, a deeper assessment of the extent of flavor conversion is still lacking.

This paper expands on our earlier work~\cite{Shalgar:2022rjj}, where we have reported on the impact of fast flavor conversion on the decoupling of neutrinos from matter in core-collapse supernovae. Our paper begins in Sec.~\ref{formulation}, where we introduce the formalism and outline the setup of our model. In Sec.~\ref{sec:noconv}, we compute the classical steady state distribution of neutrinos in the absence of flavor transformation. First, we explore how the angular distributions of neutrinos evolve as functions of the radius and become forward peaked as neutrinos decouple from matter. Then, for illustrative purposes, we carry out the linear stability analysis restricting ourselves to  the homogenous mode and by relying on the classical steady state distributions. Section~\ref{sec:conv}  focuses on the non-linear regime of flavor conversion in the presence of neutrino-neutrino interaction, advection, and collisions with the matter background. Finally, we discuss and summarize our findings in Sec.~\ref{sec:outlook}. Appendix~\ref{app:num} provides additional details on the numerical convergence of the simulations.

\section{Problem setup}
\label{formulation}
In this section, we  introduce the neutrino equations of motion. Then, we provide details on the simulation setup and the parameters adopted to model the neutrino flavor evolution. 
\subsection{Neutrino equations of motion}

For the sake of simplicity,  we assume that neutrinos are monoenergetic and work in the two flavor $(\nu_e,\nu_x)$ approximation. However, note that additional modifications to the flavor conversion physics may be derived by relaxing such approximations~\cite{Shalgar:2021wlj,Shalgar:2020xns,Capozzi:2022dtr,Capozzi:2020kge}. 

\begin{figure}[b]
\includegraphics[width=0.49\textwidth]{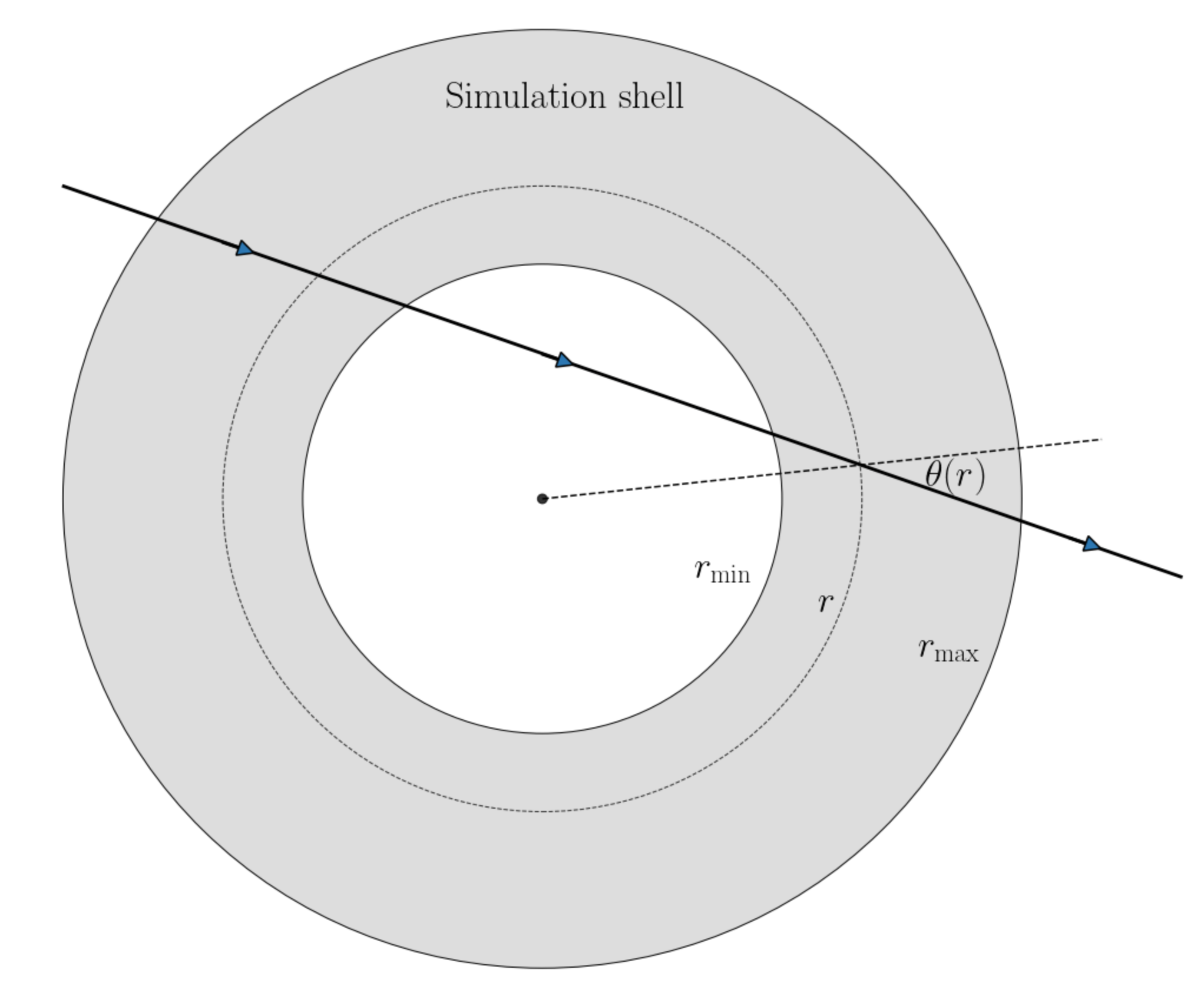}
\caption{Schematic diagram of our simulation shell. The gray shell illustrates the geometry of the region in which the simulation is carried out from $r_{\mathrm{min}}$ to $r_{\mathrm{max}}$. The dotted line shows the radial direction, while the solid line represents a straight neutrino trajectory. Neutrinos can be emitted, absorbed, or undergo direction changing interactions along any  trajectory. For any given trajectory, the  angle with respect to the radial direction [$\theta(r)$]  depends on the radius. Note that, because of spherical symmetry, the neutrino angular distributions are identical irrespective of the orientation of the radial direction for all flavors.}
\label{cartoon}
\end{figure}

The evolution of flavor  can be  investigated in terms of Wigner transformed $2 \times 2$ density matrices in the flavor space for neutrinos and 
antineutrinos,
$\rho(r,\cos\theta,t)$ and $\bar{\rho}(r,\cos\theta,t)$, respectively.  The diagonal elements of the density matrix, $\rho_{ii}$ (with $i=e, x$), stand for the occupation numbers of neutrinos of different species, while the off-diagonal terms $\rho_{ij}$ encode flavor coherence.
As shown in Fig.~\ref{cartoon}, the parameter $r$ represents the radial direction, while $\theta \equiv \theta(r)$ is the angle with respect to the radial direction at the given $r$, which should not be confused with the emission angle. The time variable is represented by $t$. 

The equations of motion that determine the flavor evolution of neutrinos and antineutrinos are given by~\cite{Sigl:1993ctk}:
\begin{eqnarray}
\label{eoms1}
i\left(\frac{\partial}{\partial t} + \vec{v}\cdot \nabla \right) \rho(r,\cos\theta,t) &=& [H,\rho(r,\cos\theta,t)] + i\mathcal{C}\\  
i\left(\frac{\partial}{\partial t} + \vec{v}\cdot \nabla \right) \bar{\rho}(r,\cos\theta,t) &=& [\bar{H}, \bar\rho(r,\cos\theta,t)] + i\bar{\mathcal{C}}\ .
\label{eoms}
\end{eqnarray}
The left-hand sides of Eqs.~\ref{eoms1} and \ref{eoms} contain the total derivative, including the advective term. 
 Due to the radial dependence of $\theta$ and the assumption of spherical symmetry, the advective term can be written as follows~\cite{Rampp:2002bq}:
\begin{eqnarray}
& &\vec{v} \cdot  \nabla \rho(r,\cos\theta,t) = \cos\theta \frac{d}{dr} \rho(r,\cos\theta,t) \nonumber \\
&=& \cos\theta \frac{\partial \rho(r,\cos\theta,t)}{\partial \cos\theta} \frac{d\cos\theta}{dr} 
+ \cos\theta \frac{\partial \rho(r,\cos\theta,t)}{\partial r}\nonumber \\
&=& \frac{\partial \rho(r,\cos\theta,t)}{\partial \cos\theta} \frac{\sin^{2}\theta}{r} + \cos\theta \frac{\partial \rho(r,\cos\theta,t)}{\partial r} \ .
\label{advective}
\end{eqnarray} 

The right-hand sides of Eqs.~\ref{eoms1} and \ref{eoms} consist of the Hamiltonian that governs the flavor evolution and the collision term. 
The Hamiltonian   includes the vacuum and  self-interaction terms:
\begin{eqnarray}
\label{Ham}
H = H_{\textrm{vac}} + H_{\nu\nu}\ ,
\end{eqnarray}
with
\begin{eqnarray}
\label{Hvac}
&&H_{\textrm{vac}} = \frac{\omega}{2}
\begin{pmatrix}
- \cos 2 \vartheta_{\textrm{V}} & \sin 2 \vartheta_{\textrm{V}} \cr
\sin 2 \vartheta_{\textrm{V}} & \cos 2 \vartheta_{\textrm{V}} 
\end{pmatrix}\\ 
&&H_{\nu\nu} = \mu_{\textrm{0}} \int  [\rho(\cos\theta^\prime)-\bar{\rho}(\cos\theta^\prime)]
\times  (1-\cos\theta\cos\theta^{\prime}) d\cos\theta^{\prime}\ . \nonumber\\
\label{Hself}
\end{eqnarray}
We use $\vartheta_{\textrm{V}}$ to denote the vacuum mixing angle, while $\omega = {\Delta m^{2}}/{2 E}$ is the vacuum frequency with $E$ being the neutrino energy.   In the self-interaction Hamiltonian, $H_{\nu\nu}$, 
$\mu_{0}$  denotes the self-interaction strength.
 The additional integration over the azimuthal angle in Eq.~\ref{Hself} results in a factor $2\pi$, which has been absorbed in  $\mu_{0}$. Note that, due to the radial evolution of the angular distributions of neutrinos,  
the effective self-interaction strength decreases as a function of the radius in the regions beyond the neutrinosphere. 
The Hamiltonian  governing the evolution of antineutrinos is the same as Eq.~\ref{Ham}, with $H_{\textrm{vac}} \rightarrow -H_{\textrm{vac}}$. The matter term in the Hamiltonian is neglected since its effect  is to reduce the effective mixing angle of neutrinos~\cite{Esteban-Pretel:2008ovd}.

\setlength\extrarowheight{1pt}
\begin{table*}
\caption{Length scales associated with emission, absorption, and direction changing scattering for Cases A, B and C. The collision terms for $\nu_{e}$ and $\nu_{x}$ are identical for all cases, but they differ from each other for $\bar{\nu}_{e}$. The function $\xi(r)$ is used to represent the function $\exp(15-r)$. All the collision terms are in units for km$^{-1}$ and $r$ is in km. 
}
\label{Tab1}
\begin{tabular}{|l|l|l|l|l|l|}
\hline
& \ \ \ \ \ \ \ \ \ \ $\nu_{e}$  &\ \ \ \ \ \  $\bar{\nu}_{e}$  &\ \ \ \ \ \   $\bar{\nu}_{e}$  &\ \ \ \ \ \   $\bar{\nu}_{e}$  &\ \ \ \ \ \ \  $\nu_{x}, \bar\nu_x$  \cr
& (Cases A, B, C) &\    (Case A) &\    (Case B) &\  (Case C) &  (Cases A, B, C) \cr
\hline
\hline
$\lambda_{\textrm{emission}}^{\nu_i}$ (km) &\ \ \ \ 1/[50 $\xi(r)]$ & 1/[50 $\xi(r)]$ & 1/[26 $\xi(r)]$ & 1/[30 $\xi(r)]$ &\ \ \  1/[10 $\xi(r)]$ \cr
$\lambda_{\textrm{absorb}}^{\nu_i}$ (km) &\ \ \ \  1/[50 $\xi(r)]$ & 1/[50 $\xi(r)]$ & 1/[25 $\xi(r)]$ & 1/[25 $\xi(r)]$ &\ \ \   1/[10 $\xi(r)]$ \cr 
$\lambda_{\textrm{dir-ch}}^{\nu_i}$ (km) &\ \ \ \  1/[50 $\xi(r)]$ & 1/[25 $\xi(r)]$ & 1/[25 $\xi(r)]$ & 1/[25 $\xi(r)]$ &\ \ \   1/[12.5 $\xi(r)]$ \cr
\hline
\end{tabular}
\end{table*}
  The collision term includes  emission, absorption, and direction-changing collisions, respectively~\cite{weinberg_2019}, i.e.~it takes into account the main reactions of neutrinos with the matter background~\cite{Shalgar:2019kzy,1982ApJS...50..115B,OConnor:2014sgn,Mezzacappa:2020oyq,Richers:2019grc}. This implies $\mathcal{C} \equiv \mathcal{C}(\vec{r}, E, t) =  \cemit + \cabsorb + \cdirch$:
\begin{eqnarray}
\label{cemission}
\cemit^{\nu_{e},\bar{\nu}_{e},\nu_{x},\bar\nu_x} &=& \frac{1}{\lambda_{\textrm{emission}}^{\nu_{e},\bar{\nu}_{e},\nu_{x},\bar\nu_x}(r)} \ , \\
\label{cabsorb}
\cabsorb^{\nu_{e},\bar{\nu}_{e},\nu_{x},\bar\nu_x} &=& - \frac{1}{\lambda_{\textrm{absorb}}^{\nu_{e},\bar{\nu}_{e},\nu_{x},\bar{\nu}_{x}}(r)} \rho_{ii}(\cos\theta)\ , \\
\label{cdirch}
\cdirch^{\nu_{e},\bar{\nu}_{e},\nu_{x},\bar\nu_x} &=& - \frac{2}{\lambda_{\textrm{dir-ch}}^{\nu_{e},\bar{\nu}_{e},\nu_{x},\bar\nu_x}(r)} \rho_{ii}(\cos\theta) \nonumber\\
&+& \int_{-1}^{1} \frac{1}{\lambda_{\textrm{dir-ch}}^{\nu_{e},\bar{\nu}_{e},\nu_{x},\bar\nu_x}(r)} \rho_{ii}(\cos\theta^{\prime}) d\cos\theta^{\prime}\ . 
\label{collform}
\end{eqnarray}
Each of the above equations refers to all flavors as denoted by the superscripts. 
 In principle, $\mathcal{C} \equiv \mathcal{C}(r,\cos\theta, E, t)$ and the  ratio among the different terms entering the collision term changes as a function of energy and time~\cite{1982ApJS...50..115B,OConnor:2014sgn,Mezzacappa:2020oyq,Richers:2019grc}. However, for the sake of simplicity, we omit any dependence on  $E$ and $t$. 
In addition, due to the small time scales associated with fast flavor evolution compared to the collision term, the off-diagonal components do not play any significant role (we have numerically verified that this assumption holds; results not shown here).
We also neglect the Pauli blocking and neutrino chemical potentials for the sake of simplicity, although they should be taken into account once more a advanced modeling of the collision term is developed~\cite{1985ApJS...58..771B,Raffelt:1996wa}.

\subsection{Simulation setup}
We carry out the simulations presented in this paper in a ``simulation shell,'' see gray shaded region in Fig~\ref{cartoon}. The radial range extends from  $r_{\textrm{min}} = 15$~km to $r_{\textrm{max}} = 30$~km, while  $\cos\theta(r) \in [-1,1]$ at each  $r$. We use a grid of $150$ uniform bins for both $\cos\theta$ and $r$. We have tested the convergence of the code with respect to the number of bins and provide further details in Appendix~\ref{app:num}.
We use $E= 20$~MeV $\Delta m^{2} = 2.5 \times 10^{-3}$ eV$^{2}$, $\mu_0 = 10^{4}$ km$^{-1}$, and the effective vacuum mixing angle  
$\vartheta_{\textrm{V}}=10^{-3}.$ 

At $r=r_{\textrm{min}}$, the boundary condition is determined by the collision term. At $r=r_{\textrm{max}}$, we impose two different boundary conditions depending on $\cos\theta$. For $\cos\theta>0$,  neutrinos  stream outward and, hence, the boundary condition is determined by conditions within the simulation region. For $\cos\theta \leq 0$, we impose a vanishing boundary condition.

In order to investigate the dependence of fast flavor conversion on the shape of the ELN crossings, we consider three different collision terms: Cases A, B, and C, 
 engineered  to give different types of ELN crossings. Case A is also adopted in Ref.~\cite{Shalgar:2022rjj}.
 The flavor-dependent length-scales entering the collision terms in Eqs.~\ref{cemission}--\ref{collform} are reported in Table~\ref{Tab1}. For all  collision terms, we use a simplified radial dependence defined by $\lambda_{\rm{emission, absorb, dir-ch}}^{\nu_i} \sim 1/ \xi(r)$, with $\xi(r)= \exp({15-r}/{\textrm{km}})$. We refer the reader to Appendix A of Ref.~\cite{Shalgar:2022rjj} for further details on the modeling of our heuristic collision term. 
 We parametrize  $\mathcal{C}$ to have a characteristic length scale of $\mathcal{O}(10$--$100)$~m at $r_{\rm min}$ for all cases
 and so that $\mathcal{C}$ falls exponentially as a function of $r$. We stress that this is a simplification, not aiming to reproduce realistic conditions in the supernova core, but rather allowing to pass from isotropic to forward peaked distribution within the simulation shell and to generate an ELN crossing, as discussed in the next section. Since we populate the simulation shell through collisions, the collisions term for $\nu_{e}$ in Table~\ref{Tab1} is chosen such that $r_{\rm min}$  is within the trapping region and the neutrino number density there is governed only by the ratio of $\cemit$ and $\cabsorb$.

The collision term involves factors that give rise to exponentially growing and damping solutions, which make them stiff. We use the Adams–Bashforth-Moulton method from the {\tt Differentialequations.jl} package of {\tt Julia}  to solve the equations of motion~\cite{rackauckas2017differentialequations, Julia-2017}. 
Each simulation took $\mathcal{O}(2000)$ CPU hours employing shared memory on the High Performance Computing Centre at the University of Copenhagen.

\section{Classical steady state configuration: no flavor conversion}
\label{sec:noconv}
In this section, we present our results on the classical steady state configuration achieved in the absence of flavor conversion. On the basis of these findings, we then introduce the linear stability analysis to investigate the regions in the simulation shell where the development of flavor instabilities is foreseen. 

\subsection{Angular distributions of neutrinos}
\label{sec:ang_distrib}
In the absence of flavor conversion (i.e.,~$H=\bar{H}=0$ in Eqs.~\ref{eoms1} and \ref{eoms}), we aim to find a classical steady state configuration by setting $\rho_{ii}=0$ as the initial condition for all flavors (the off-diagonal terms of the density matrices $\rho_{ij}$ are initially equal to zero and remain as such throughout the evolution in the absence of flavor transformation) and by considering the collision and advection terms only. 

Neutrinos in our setup are generated through collisions and advected across the simulation shell.
The advective term allows for a change in the number density of neutrinos at a given location due to their motion. 
As for the collision term,  the emission term is independent of the number density of neutrinos,  the absorption term is proportional to the number density of neutrinos, while the direction changing term conserves the number of neutrinos. 
To obtain the steady state configuration, we need to evolve the neutrino field in our simulation shell at least for a period corresponding to the radial range $[r_{\rm min}, r_{\rm max}]$,  assuming that neutrinos travel at the speed of light, i.e.~$t=5\times 10^{-5}$~s. In the simulations, we have evolved the system for $t=10^{-4}$~s out of caution.

\begin{figure}
\includegraphics[width=0.49\textwidth]{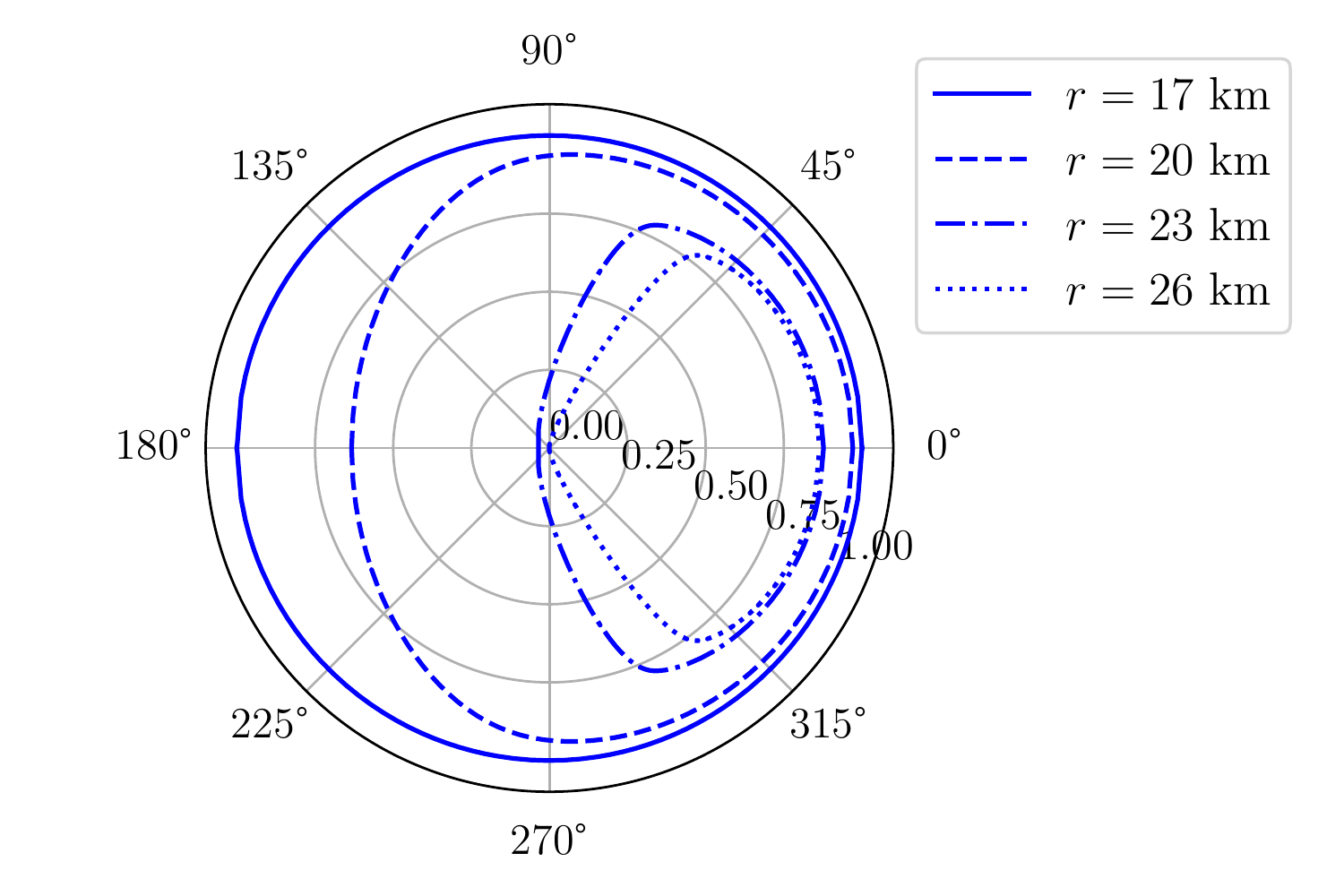}\\
\hspace{-.8cm}\includegraphics[width=0.49\textwidth]{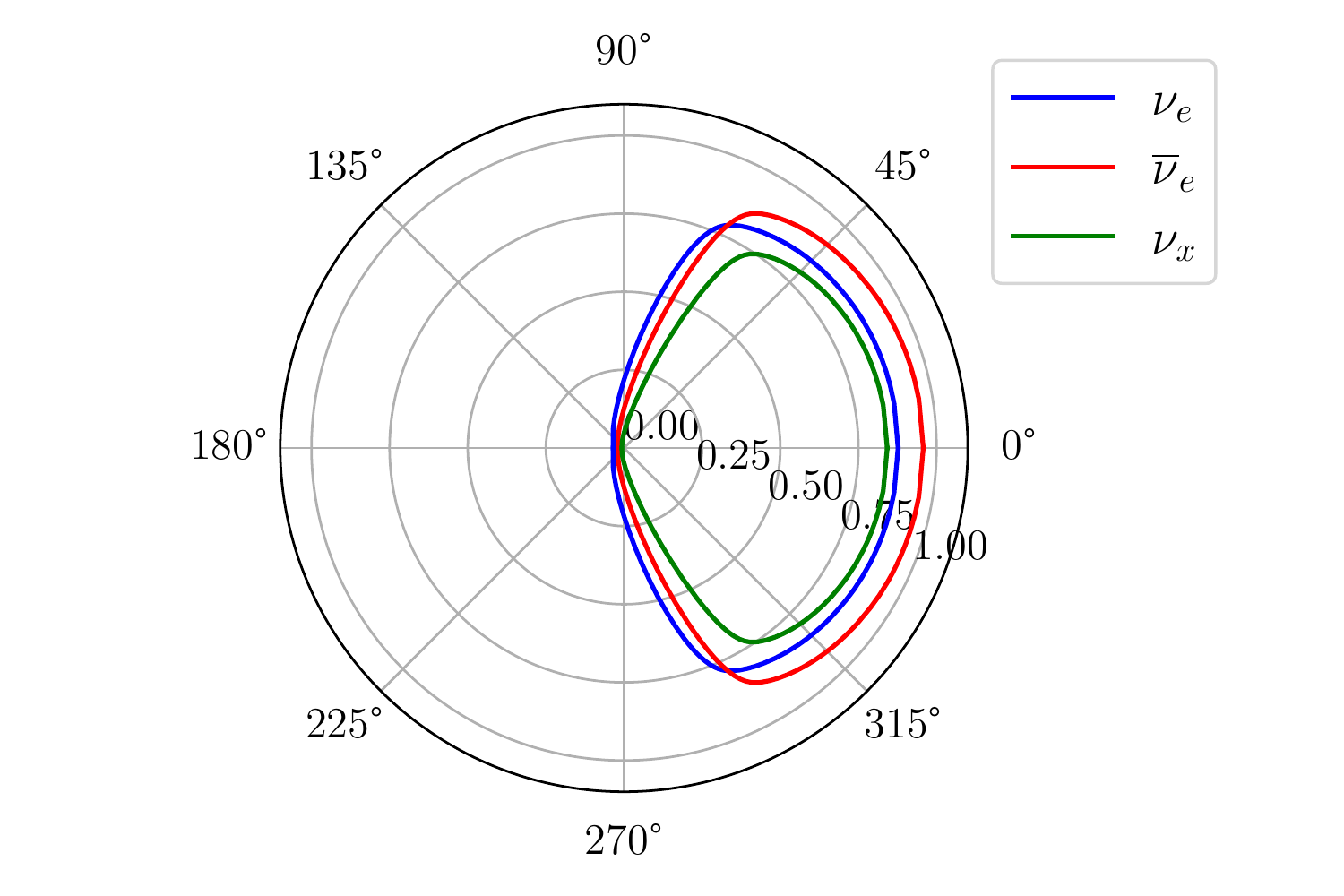}
\caption{{\it Top:} Polar diagram of the steady state angular distribution of $\nu_e$ for different radii for Case C in the absence of neutrino transformation. The angle $\theta=0$ corresponds to the local radial direction. At small radii  the angular distribution is isotropic and slowly becomes forward peaked with falling matter density. {\it Bottom:}  Polar diagram of the angular distributions of $\nu_e$, $\bar\nu_e$ and $\nu_x$ at $r=23$~km. Because of the different interaction rates with matter, the distributions of neutrinos of different flavors are not equally forward peaked, with the angular spread being the largest for $\nu_e$, followed by $\bar\nu_e$ and $\nu_x$.} 
\label{Fig2}
\end{figure}
The top panel of Fig.~\ref{Fig2} shows a polar diagram of the angular distribution of $\nu_e$ for Case C, which we consider  our benchmark configuration hereafter, once the steady state configuration is achieved. We obtain an isotropic configuration at small radii, which slowly becomes forward peaked at larger radii as the density falls. Such a trend holds for all flavors. However, as we move towards larger radii and the density falls, $\nu_x$ start forward peaking, followed by $\bar\nu_e$ and $\nu_e$, as displayed in the bottom panel of Fig.~\ref{Fig2}. This behavior can lead to ELN crossings and hence fast flavor instabilities.
The qualitative trend in the angular distributions for Cases A and B is similar to the one of Case C and therefore not shown here.

The classical steady state obtained as described above constitutes the initial configuration adopted to solve the neutrino equations of motion including flavor conversion (see Sec.~\ref{sec:conv}). This procedure is important from a numerical point of view. Any configuration that is not initially in a classical steady state leads to large gradients at the edges of the simulation shell, which give rise to numerical instabilities. Moreover, 
the advective term involves the calculation of the derivatives using a finite-element method, leading to numerical instabilities without sufficient resolution. Various tests have been carried out to make sure that numerical instabilities do not affect the results presented here.

\subsection{Looking for flavor instabilities through the  classical steady state solutions}

\begin{figure}
\includegraphics[width=0.49\textwidth]{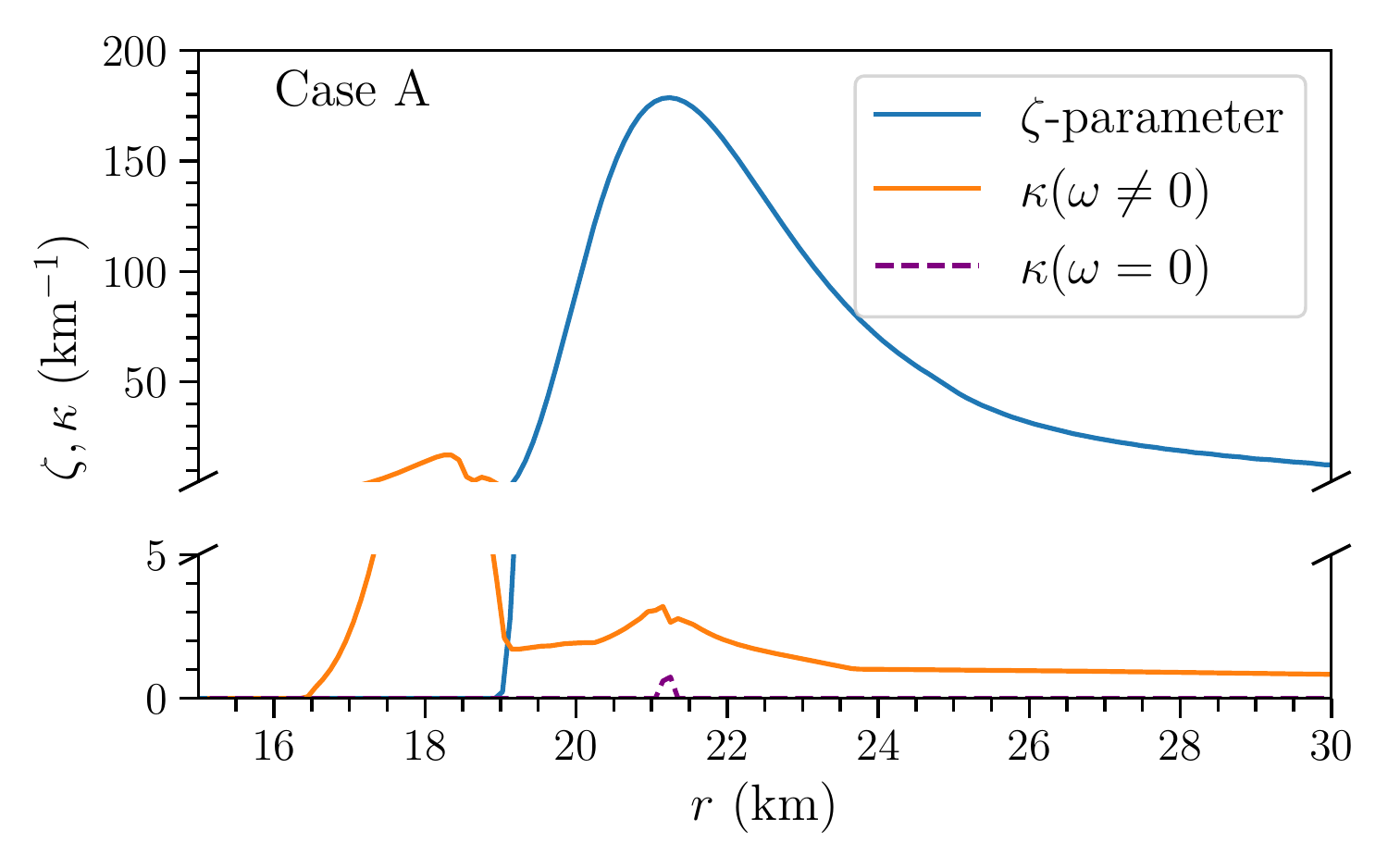}
\includegraphics[width=0.49\textwidth]{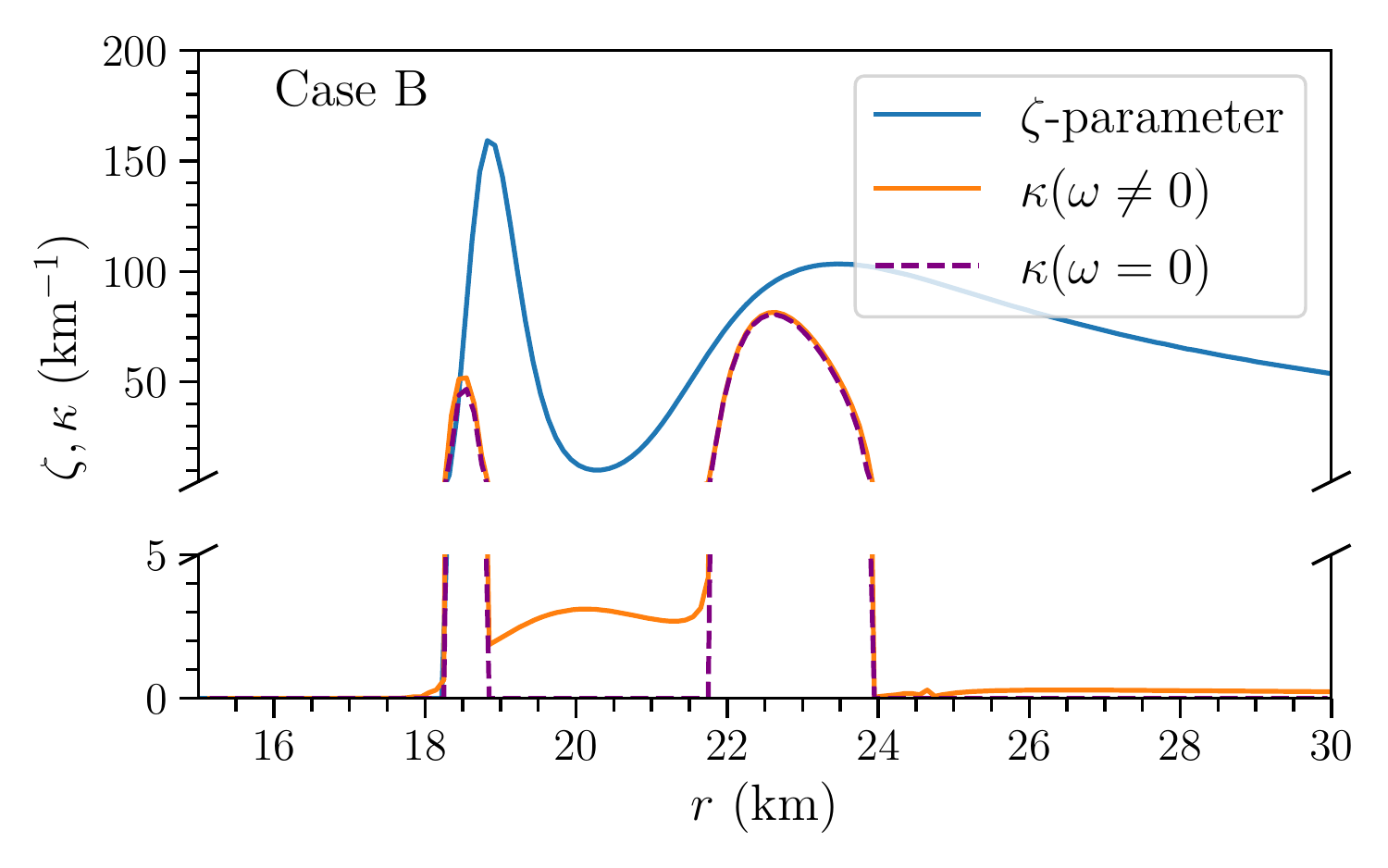}
\includegraphics[width=0.49\textwidth]{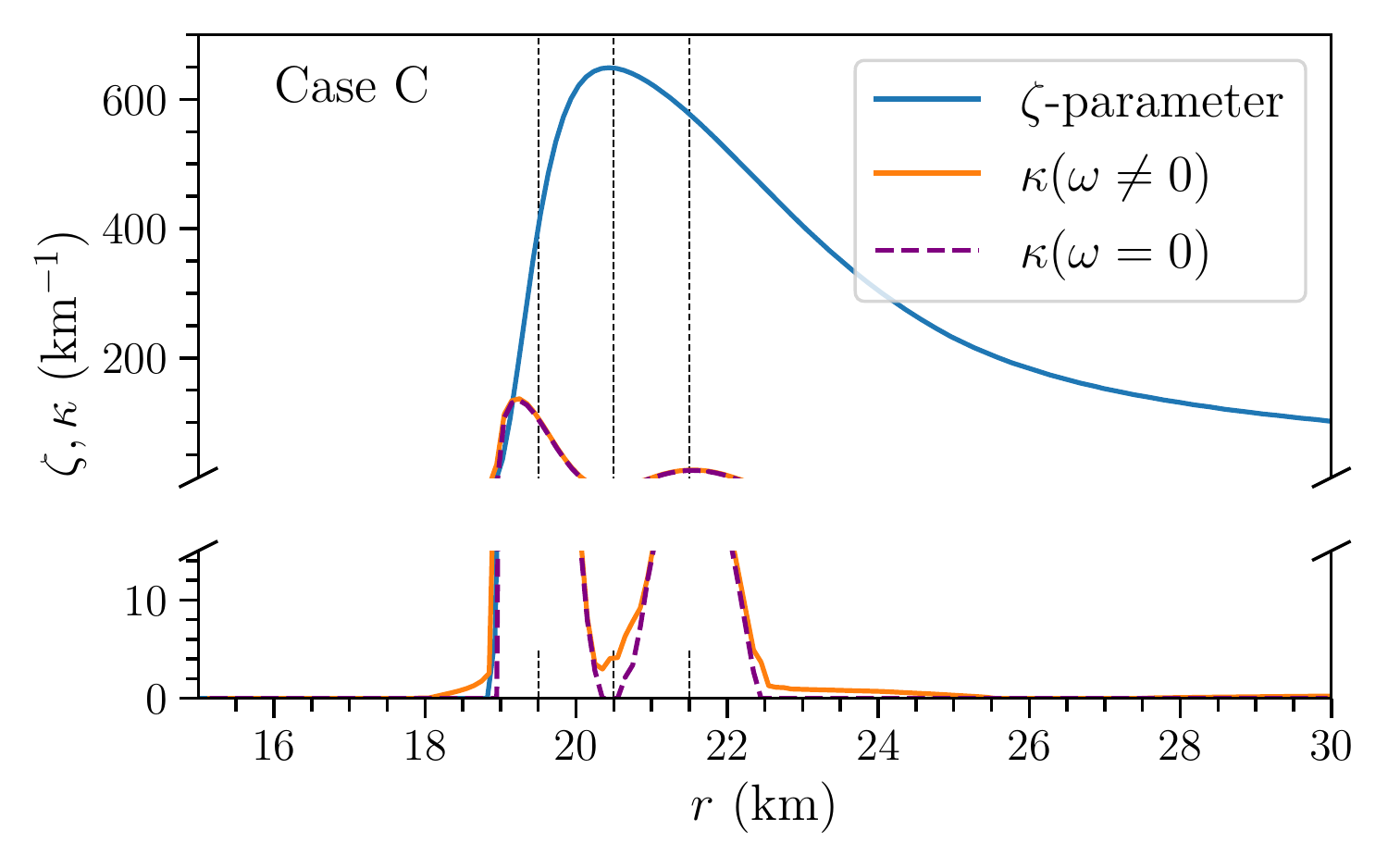}
\caption{Radial profiles of the $\zeta$ parameter (see Eq.~\ref{eq:zeta}; solid blue line) and growth rates calculated for the homogeneous mode using the linear stability analysis for the cases of vanishing (dashed magenta line) and non-vanishing (solid orange line) vacuum frequency for 
Cases A, B, and C from top to bottom, respectively. The non-zero $\zeta$-parameter implies the existence of an ELN crossing, which is a necessary condition for the existence of fast flavor instability. In all three cases, the $\zeta$ parameter peaks where the $\omega = 0$ flavor instability occurs. Moreover, there are radii for which a flavor instability exists only for $\omega \neq 0$. A broken $y$-axis is adopted because of the significant difference in the growth rates of the instability for the fast ($\omega=0$) and the slow ($\omega\not=0$) cases. The difference between the orange solid line and the purple dashed line highlights  the impact of the vacuum frequency on the growth rate. For radii where the fast growth rate is substantial, the vacuum term has negligible effect. But when the fast growth rate is small or absent the vacuum term can become important.
The dashed vertical lines in the bottom panel denote the radii at which the absolute values of the components of the eigenvectors are shown in Fig.~\ref{eigenvectors}.
}
\label{growth}
\end{figure}

It has been proven that the existence of ELN crossings is a necessary condition for fast flavor instabilities~\cite{Morinaga:2021vmc,Izaguirre:2016gsx}. In order to gauge the presence of ELN crossings, we rely on a slightly modified definition of  the $\zeta$ parameter introduced in Ref.~\cite{Padilla-Gay:2020uxa} and evaluate it at the time when the steady state configuration has been reached:
\begin{eqnarray}
\label{eq:zeta}
\zeta(r) = \mu_{0} \frac{I_{1}(r) I_{2}(r)}{I_{1}(r)+I_{2}(r)}\ ,
\end{eqnarray}
with
\begin{eqnarray}
I_{1}(r) = \int [\rho_{ee}(r, \cos\theta)-\bar{\rho}_{ee}(r, \cos\theta)] d\cos\theta
\end{eqnarray} 
for  $\rho_{ee}(r, \cos\theta) > \bar{\rho}_{ee}(r, \cos\theta)$ and
\begin{eqnarray}
I_{2}(r) = \int [\bar{\rho}_{ee}(r, \cos\theta)-\rho_{ee}(r, \cos\theta)] d\cos\theta 
\end{eqnarray}
for  $\rho_{ee}(r, \cos\theta) < \bar{\rho}_{ee}(r, \cos\theta)$.
The $\zeta$ parameter can be different from zero if and only if there are  regions in the angular domain where $\rho_{ee}(\cos\theta) > \bar\rho_{ee}(\cos\theta)$ and other regions in the angular domain where $\rho_{ee}(\cos\theta) < \bar\rho_{ee}(\cos\theta)$, which implies the existence of an ELN crossing. 
Figure~\ref{growth} shows the radial profile of  $\zeta$ for Cases A, B, and C. An ELN crossing exists for $r \gtrsim 18$ km for all cases; hence one should expect a potential flavor instability in this region.

To better gauge which regions of the simulation shell may be prone to flavor instabilities, we rely on the linear stability analysis of the classical steady state solution obtained in Sec.~\ref{sec:ang_distrib} for Cases A, B, and C (see Tab.~\ref{Tab1}). In particular, we focus on flavor instabilities in the limit of vanishing and non-vanishing vacuum frequency (i.e., fast and slow flavor instabilities). For the sake of simplicity, we focus on the  linear stability analysis  for the homogeneous mode only, since we aim to gain insight on where flavor conversion may develop. This simplifying choice is also justified by the fact that if the neutrino gas is not homogeneous, as in our case,  the equations for the  Fourier modes are coupled. Note, however, that the numerical results presented in Sec.~\ref{sec:conv} do not distinguish between homogeneous and inhomogeneous modes and do take into account collisions and advection.   

To this purpose, we  linearize Eqs.~\ref{eoms1} and \ref{eoms} ignoring the collision and  advective terms. The linearization implies expanding the equations of motion for the off-diagonal components of the density matrix $\rho_{ex}(r,\cos\theta,t)$ up to linear order in $\rho_{ex}(r,\cos\theta,t)$ (and the same for $\bar{\rho}_{ex}(r,\cos\theta,t)$)~\cite{Banerjee:2011fj, Izaguirre:2016gsx}.  For each $r$, the results are solutions for $\rho_{ex}(r,\cos\theta,t)$ and $\bar{\rho}_{ex}(r, \cos\theta,t)$ of the form:
\begin{eqnarray}
\rho_{ex}(r,\cos\theta,t) \sim \exp(-i\Omega t) \rho_{ex}(r,\cos\theta,0)\ ,\\
\bar{\rho}_{ex}(r,\cos\theta,t) \sim \exp(-i\Omega t) \bar{\rho}_{ex}(r, \cos\theta,0)\ , 
\label{linsol}
\end{eqnarray}
where $\Omega$ is the eigenvalue which is independent of $\cos\theta$ and it is the same for neutrinos and antineutrinos, due to the collective nature of the flavor evolution. The eigenvalue $\Omega$ can be obtained semi-analytically and either is  real or appears in complex-conjugate pairs. A complex $\Omega$ with a non-zero imaginary part $\kappa$ implies that  $\rho_{ex}$ and $\bar{\rho}_{ex}$ grow exponentially; this  is known as flavor instability~\cite{Banerjee:2011fj}. 

The flavor instability thus obtained can be classified as a ``fast'' flavor instability, if it exists in the limit $\omega \rightarrow 0$, and  ``slow'' flavor instability otherwise. Note that this definition of slow flavor instability is a generalization of  the one commonly adopted in the literature, invoking the existence of at least one crossing between the electron and non-electron flavors either in  energy~\cite{Raffelt:2007cb,Raffelt:2007xt,Fogli:2008pt,Fogli:2007bk, Dasgupta:2009mg} or in angle~\cite{Mirizzi:2011tu,Mirizzi:2012wp}.  The crossing would determine the development of a flavor instability while conserving the lepton number. 
As discussed later, the presence of ELN crossings (in angle) for our system of mono-energetic neutrinos is enough to guarantee the development of slow flavor instabilities for $\omega \neq 0$.
The presence of a fast flavor instability requires that an ELN crossing occurs, which means that  
 there is at least one angle for which $\rho_{ee}(\cos\theta)=\bar{\rho}_{ee}(\cos\theta)$.
On the other hand, the presence of an ELN crossing does not necessarily imply the existence of a flavor instability or large flavor conversion~\cite{Padilla-Gay:2021haz,Padilla-Gay:2020uxa}.

 Figure~\ref{growth}  shows the growth rate of the flavor instability obtained by relying on the linear stability analysis. 
 A non-zero value of the growth rate $\kappa$  denotes the regions of flavor instability for $\omega =0$ and $\omega \neq 0$, and in the absence of advection and collisions. One can see that the $\zeta$ parameter peaks in the same region where the flavor instability for $\omega =0$ is most prominent, confirming that the $\zeta$ parameter is a good indicator of the regions of instability~\cite{Padilla-Gay:2020uxa}.  
  
Figure~\ref{growth} displays a substantial radial range where the flavor instability is present for all cases. However, not all regions that exhibit a flavor instability do so due to a fast flavor instability. In some radial regions, the flavor instability is due to slow collective modes, as can be seen by comparing the orange curve with the magenta one. Moreover, for Cases B and C, the growth rates for $\omega=0$ and $\omega \neq 0$ coincide for some spatial regions.  
The growth rate for the fast flavor instability is much faster than the one of the slow flavor instability for the homogeneous mode, as seen by comparing the radial range for which only the slow flavor instability is present in Fig.~\ref{growth}  with the one where the fast instability occurs. 

The fact that there are regions where the fast instabilities occur with the growth rate being strongly influenced by the vacuum term demands for a reassessment of the distinction between fast and slow flavor instabilities. In fact, the presence of a slow flavor instability near the decoupling region has not been shown in the literature before, because most studies do not follow the evolution of the angular distributions as functions of the radius or they just focus on one of the two instabilities ($\omega =0$ or $\omega \neq 0$).
The implications of the presence of slow flavor instabilities near the decoupling region could show many more interesting results in  multi-energy calculations because of the large vacuum frequency associated with the low energy tail of the neutrino distributions and its interplay with fast modes~\cite{Shalgar:2020xns,Duan:2010bg,Mirizzi:2015eza}. 

\begin{figure}
\includegraphics[width=0.47\textwidth]{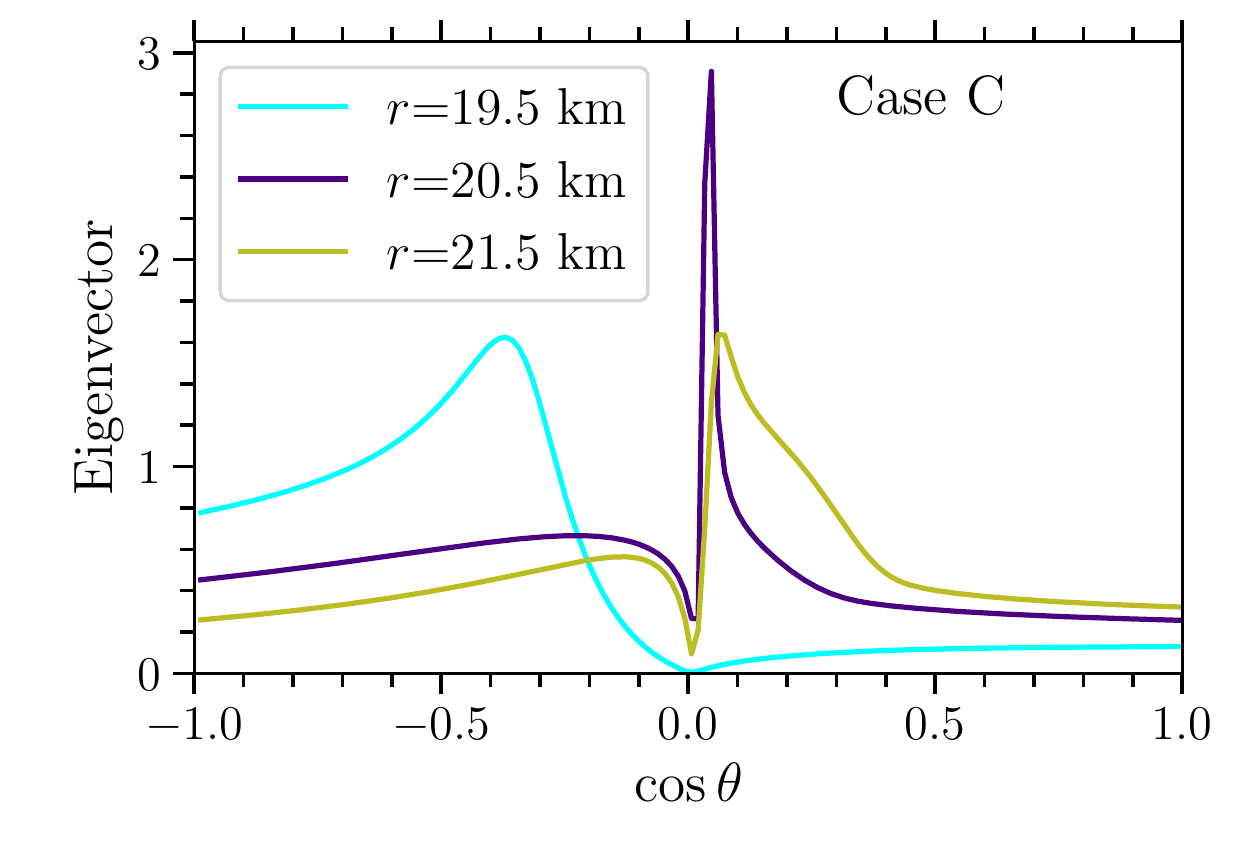}\\
\hspace{-0.4cm}\includegraphics[width=0.5\textwidth]{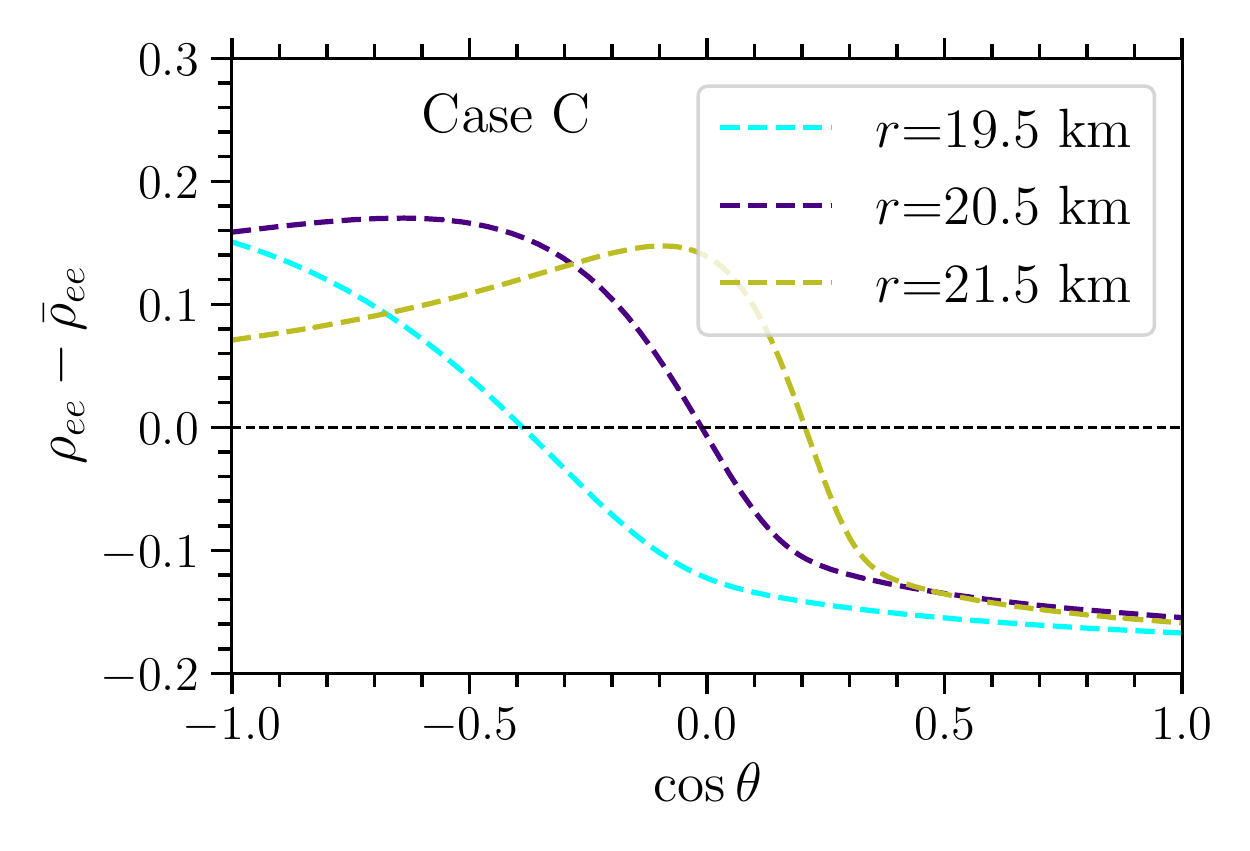}
\caption{{\it Top:} Absolute value of the components of the eigenvectors as  functions of $\cos\theta$ for Case C at $19.5$, $20.5$, and $21.5$~km  in cyan, indico, and olive,  respectively. The eigenvector components are normalized such that their integral over $\cos\theta$ is $1$. The radii at which the eigenvectors are computed are marked by vertical lines in the bottom panel of Fig.~\ref{growth}. {\it Bottom:} Difference between the  angular distributions  of $\nu_e$ and $\bar\nu_e$ in the absence of flavor transformation (i.e., ELN angular distribution)  for the same radii as in the top panel.  The eigenvector peaks in the region of the ELN crossing. 
}
\label{eigenvectors}
\end{figure}
For both slow and fast instabilities, the absolute value of the eigenvectors determines the angular regions where flavor transformation first manifests itself for the homogeneous mode. Comparing the top and bottom panels of Fig.~\ref{eigenvectors}, we can see that the absolute value of the eigenvector for Case C  peaks near the region of the ELN crossing (see also the bottom panel of Fig.~\ref{growth} where the $\zeta$ parameter peaks). It is also interesting to note that the  absolute value of the eigenvector is zero or nearly zero in the proximity of $\cos\theta=0$  (see Fig.~\ref{eigenvectors}). This divides the angular regions in two distinct regions which show qualitatively different flavor evolution, and flavor transformation in one domain ($\cos\theta \in [0,1]$) does not easily spread to the other domain ($\cos\theta \in [-1,0]$), as discussed in Sec.~\ref{sec:conv}. From Fig.~\ref{growth}, we conclude that we should expect fast instabilities for $r=19.5$ and $21.5$~km, and slow instability for  $r =20.5$~km. Interestingly, we can see from  Fig.~\ref{eigenvectors} that the slow instability for $r =20.5$~km still develops in the proximity of the ELN crossing, generalizing the findings of Refs.~\cite{Mirizzi:2011tu,Mirizzi:2012wp}.

\section{Quasi steady state configuration: flavor conversion physics}
\label{sec:conv}
In this section, we investigate the quasi steady state configuration reached by our system in the presence of flavor transformation. We then explore the effects of flavor conversion on neutrino decoupling, expanding on the findings of Ref.~\cite{Shalgar:2022rjj}, before to discuss the dynamical coupling among flavor conversion, collisions, and neutrino advection.

\subsection{Neutrino flavor transformation in the non-linear regime}  
In the presence of flavor conversion, this paper aims to find a ``quasi steady state'' solution of Eqs.~\ref{eoms1} and \ref{eoms}. In fact, due to the non-linear nature of the flavor evolution, it is not possible to obtain a flavor configuration for which the neutrino flavor remains constant as a function of time for each $\cos\theta$ and $r$. The spatial and angular structures continue to evolve on smaller and smaller scales. 
The quasi steady state configuration should be reached by solving Eqs.~\ref{eoms1} and \ref{eoms} irrespective of the initial condition. However, from a numerical perspective, it is convenient to start with a configuration that is as close to the classical steady state configuration as possible. 

\begin{figure}[]
\includegraphics[width=0.49\textwidth]{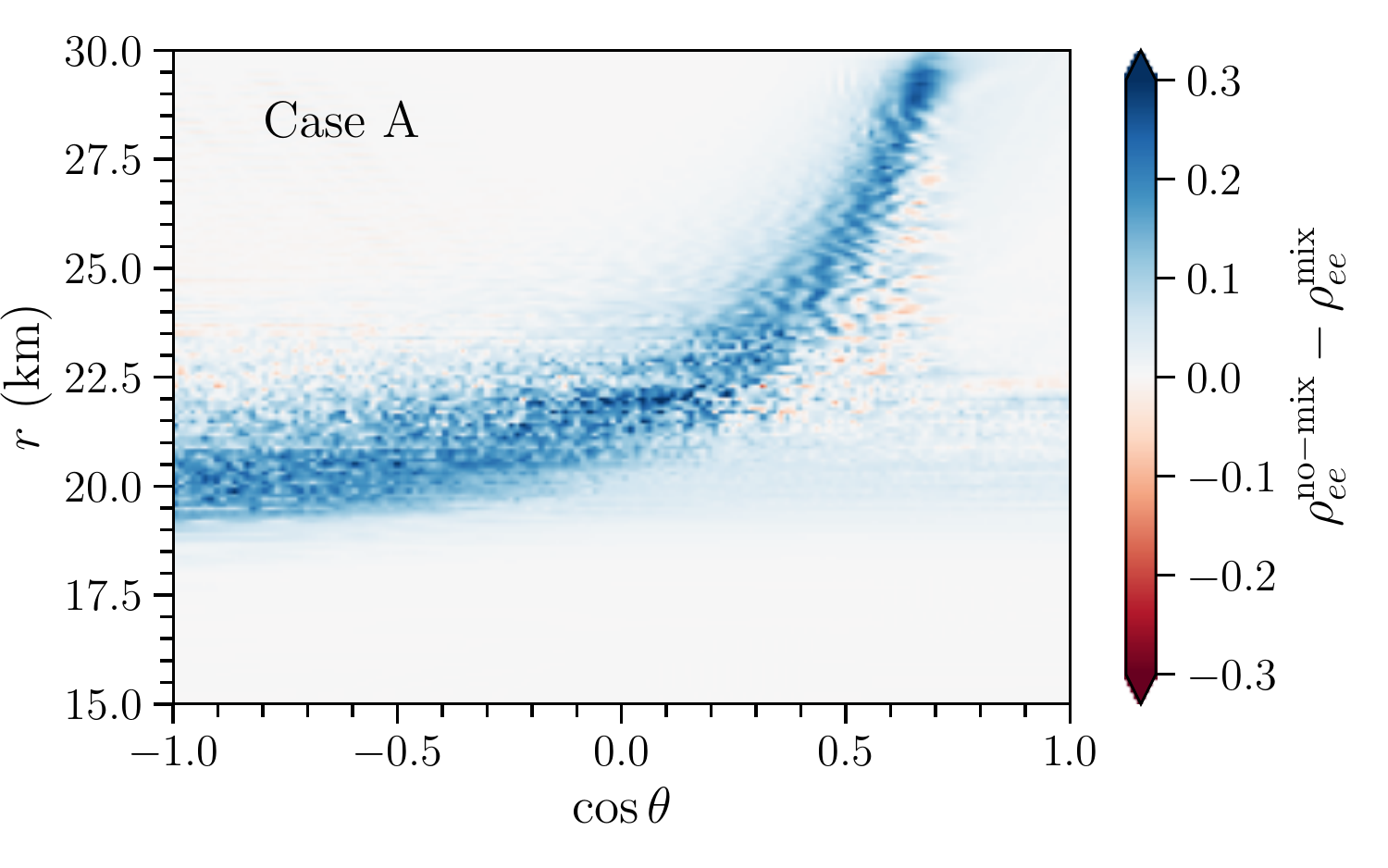}
\includegraphics[width=0.49\textwidth]{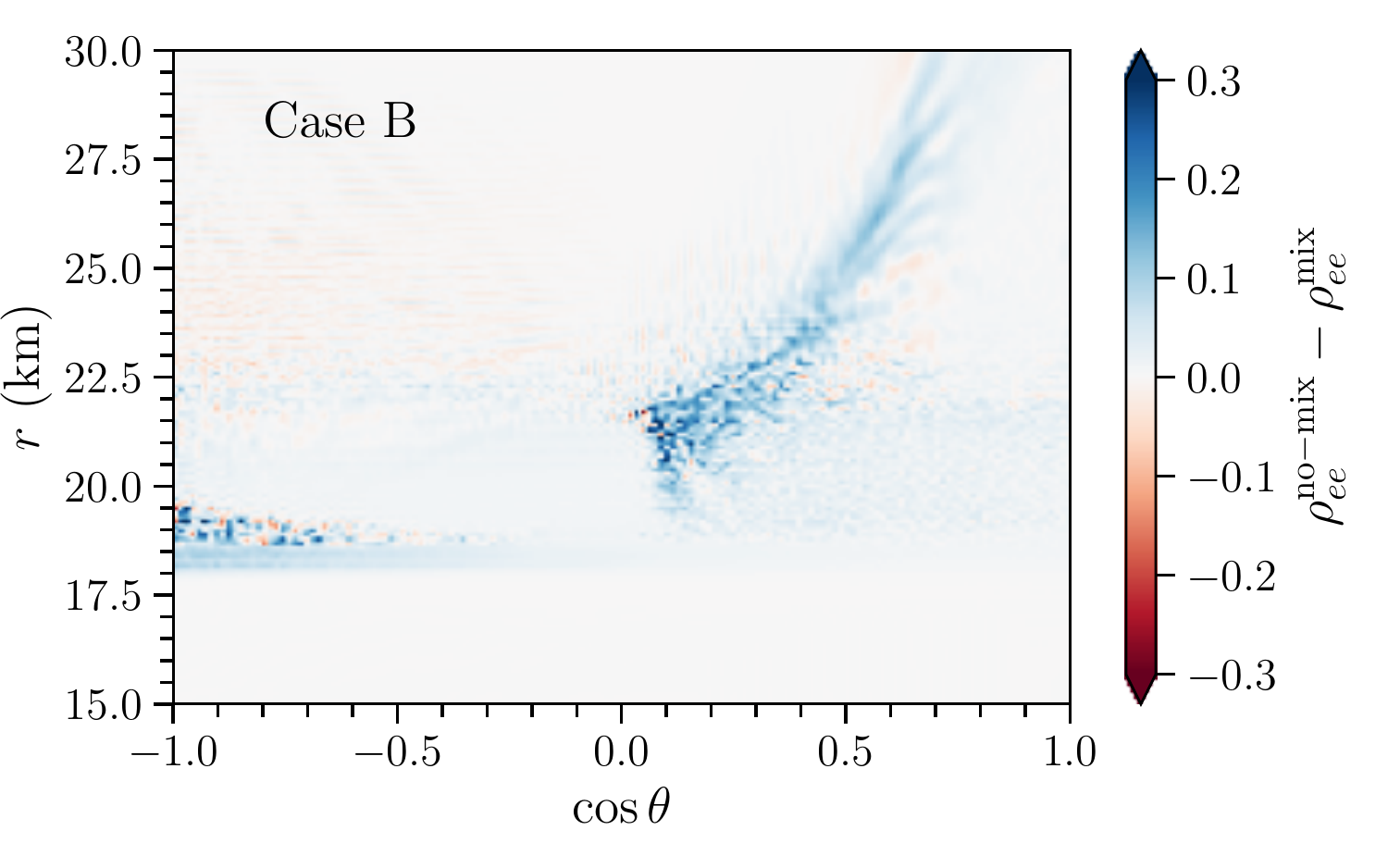}
\includegraphics[width=0.49\textwidth]{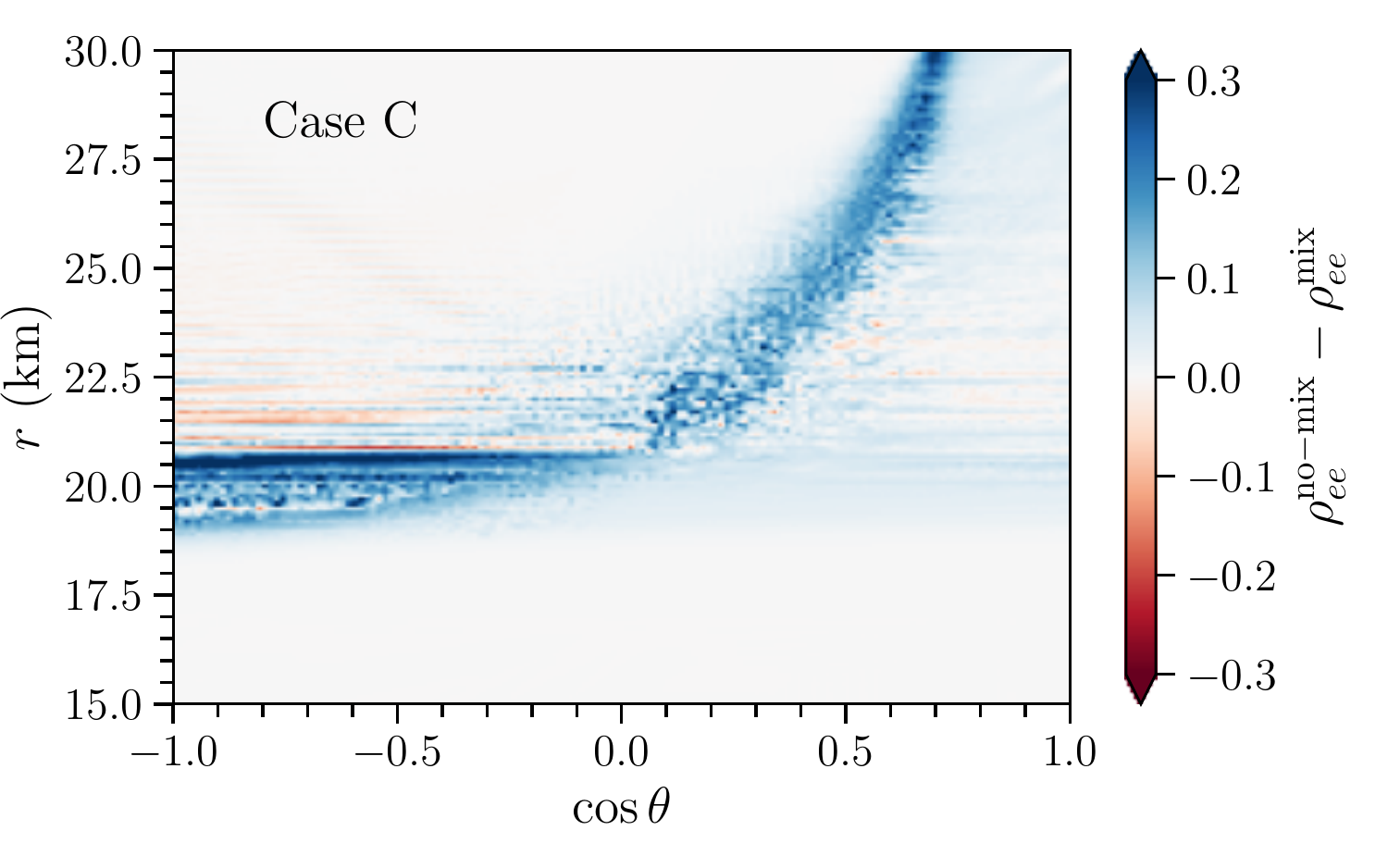}
\caption{Contour plot of the difference between  $\rho_{ee}$  with and without neutrino mixing, in the presence of collisions and advection   in the plane spanned by $\cos\theta$ and $r$ for Cases A, B and C, from top to bottom respectively. Red parts show the regions in the parameter space for which  there are less electron neutrinos due neutrino flavor transformation with respect to  the case without neutrino mixing, while the opposite is true for the blue regions. A similar trend occurs for $\bar\nu_e$ and it is not shown here.
}
\label{collandadvec}
\end{figure}
Figure~\ref{collandadvec} shows a contour plot of the difference between  $\rho_{ee}$  with and without neutrino mixing in the plane spanned by $\cos\theta$ and $r$ for Cases A, B, and C. 
The density matrix element, $\rho_{ee}^{\rm mix}$ was calculated by using the classical steady state configuration ($\rho_{ee}^{\rm no-mix}$) as the initial condition. The latter was  then evolved  up to 
$5 \times 10^{-5}$~s, which corresponds to the size of the simulation shell divided by the speed of light. 
If a smaller value of mixing angle is used, it takes longer for the system to reach the quasi steady state, but the results are unchanged.

By comparing with Figs.~\ref{collandadvec} and \ref{growth} and by taking into account the findings of Refs.~\cite{Padilla-Gay:2021haz,Shalgar:2022rjj}, we conclude that looking for flavor instabilities is not enough to predict the radial regions actually affected by flavor conversion. In fact,   the neutrinos that undergo flavor transformation at one radius are transported to larger radii due to advection. 
Cases B and C show an interesting phenomenology for the effect of advection and collisions, since the flavor instability is limited to a small range of the radial region (Fig.~\ref{growth}), but the actual region affected by flavor conversion is larger because of the dynamical effects induced by advection and collisions (Fig.~\ref{collandadvec}).

\begin{figure}
\includegraphics[width=0.49\textwidth]{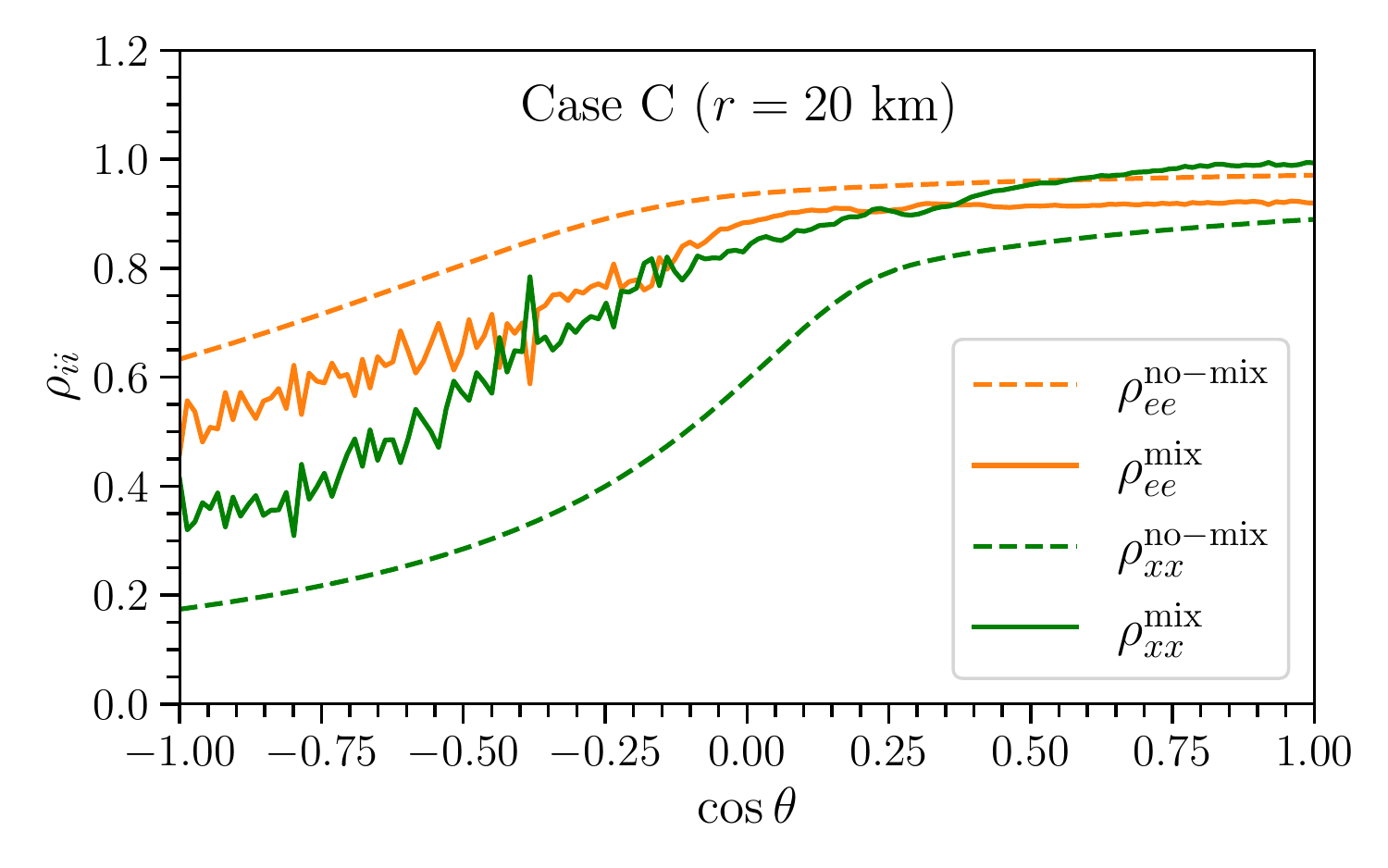}
\includegraphics[width=0.49\textwidth]{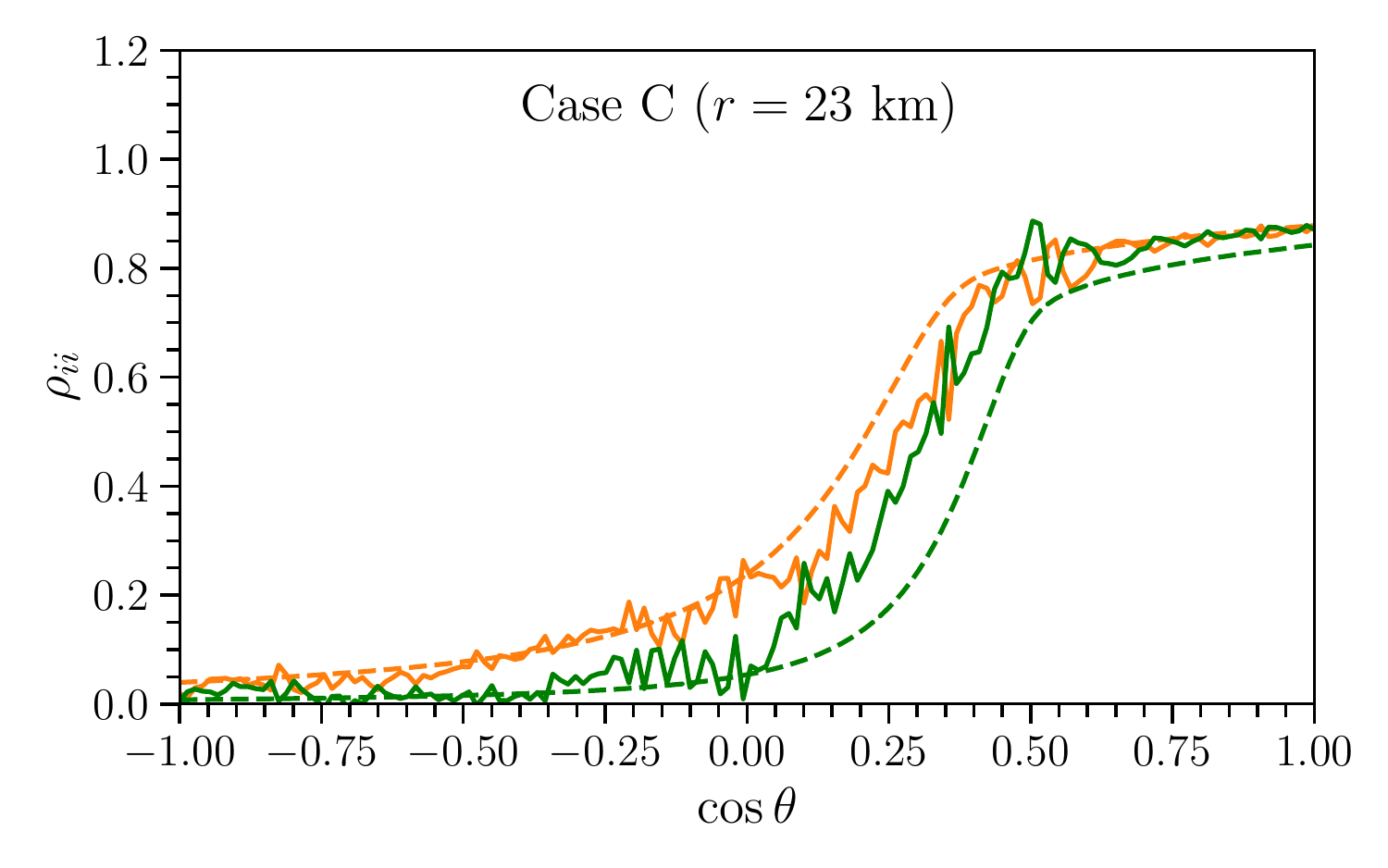}
\includegraphics[width=0.49\textwidth]{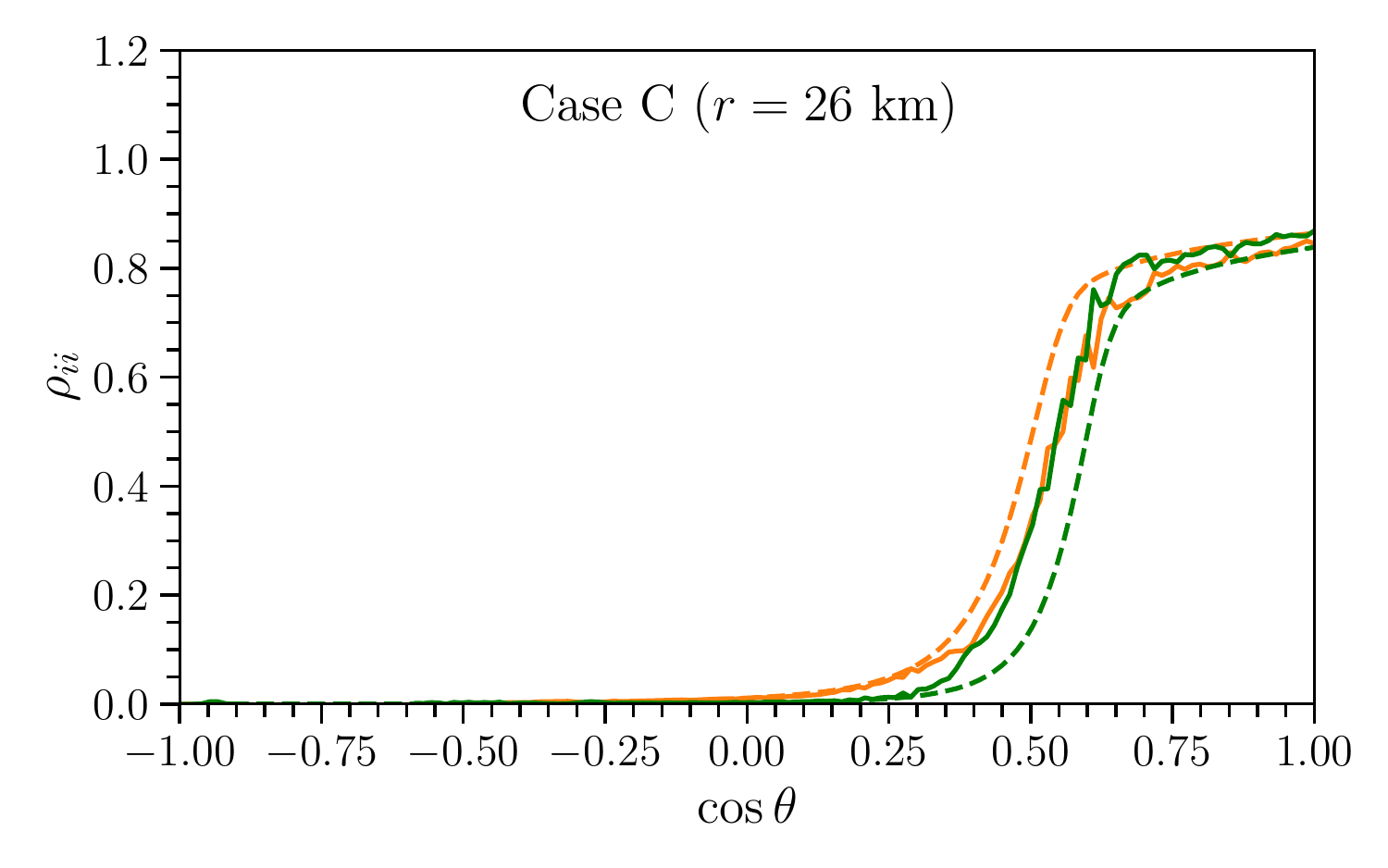}
\caption{Angular distributions of $\rho_{ee}$  (in orange) and $\rho_{xx}$ (in green)  for Case C at $20$, $23$, and $26$~km, from top to bottom, respectively. The dashed lines show the angular distributions in the absence of neutrino mixing while the solid lines show the same with neutrino mixing.  Flavor transformation affects an angular range  larger than the one where the ELN crossing is initially present. 
}
\label{angdist}
\end{figure}

The results from the numerical simulations show a spread in the flavor transformation to a finite domain of angular range, typically on one side of the ELN crossing as visible by comparing  Figs.~\ref{growth} and \ref{angdist}. To some extent, the essence of this phenomenon is captured in a homogeneous system with collisions~\cite{Shalgar:2020wcx,Tamborra:2020cul}.
From Fig.~\ref{angdist}, we can also see that, as $r$ increases, the angular distributions become more forward peaked and the distributions of $\nu_e$ and $\nu_x$ are similar to each other. As a consequence, the angular distributions after flavor transformation of  $\nu_e$ and 
$\nu_x$ tend to be similar to each other for certain $r$, however, this does not imply that flavor equipartition is a general finding.

\begin{figure}[!ht]
\includegraphics[width=0.49\textwidth]{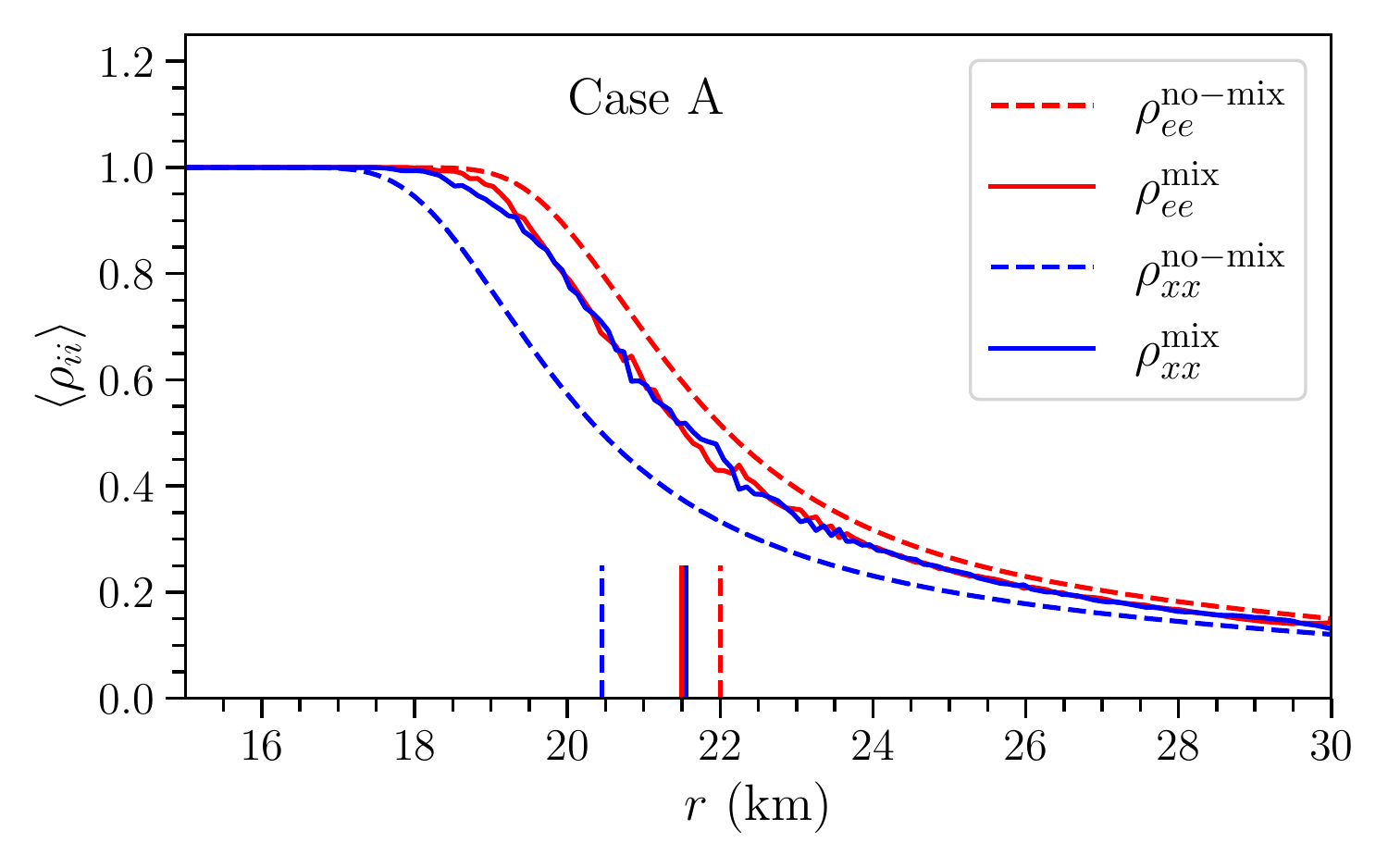}
\includegraphics[width=0.49\textwidth]{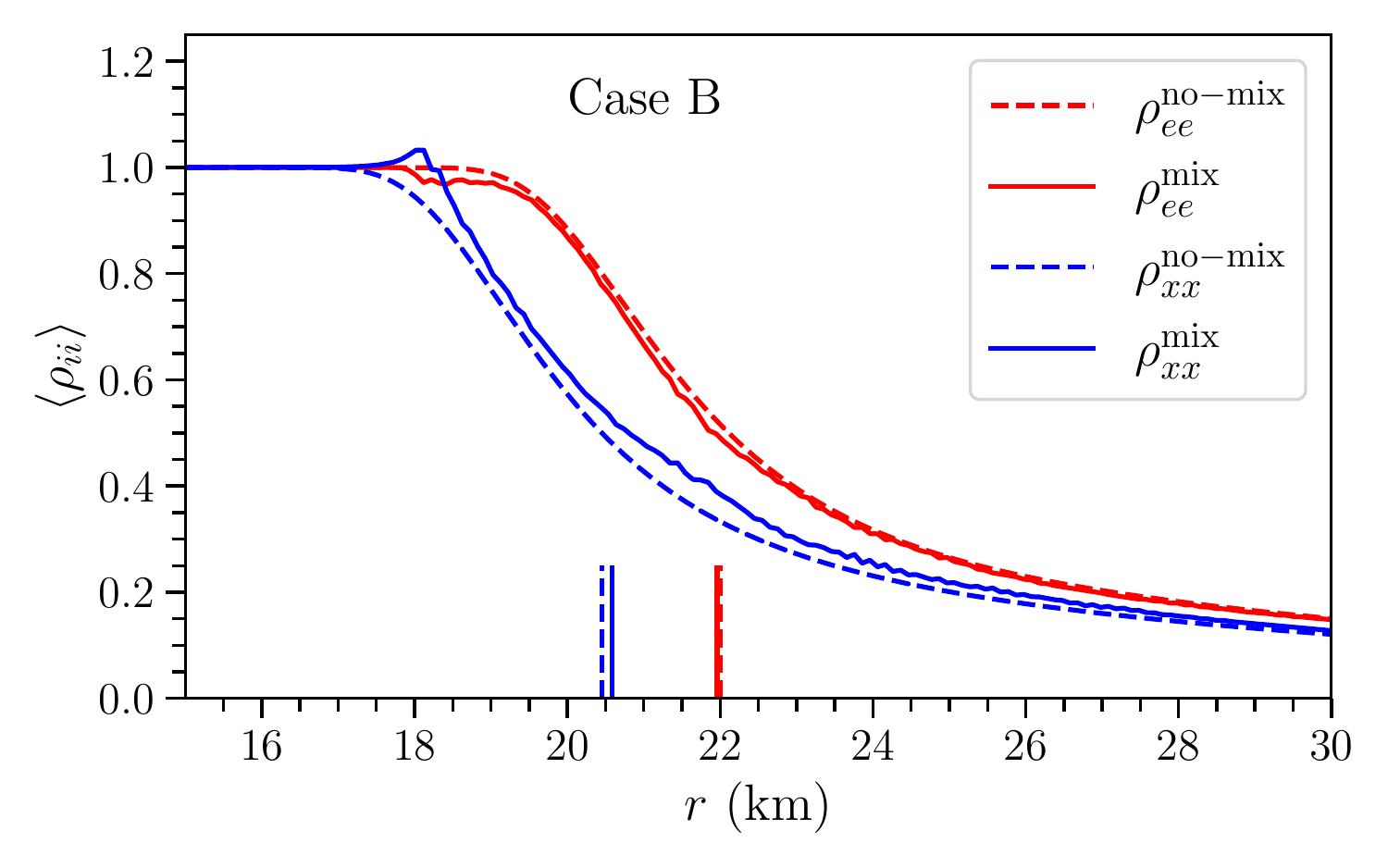}
\includegraphics[width=0.49\textwidth]{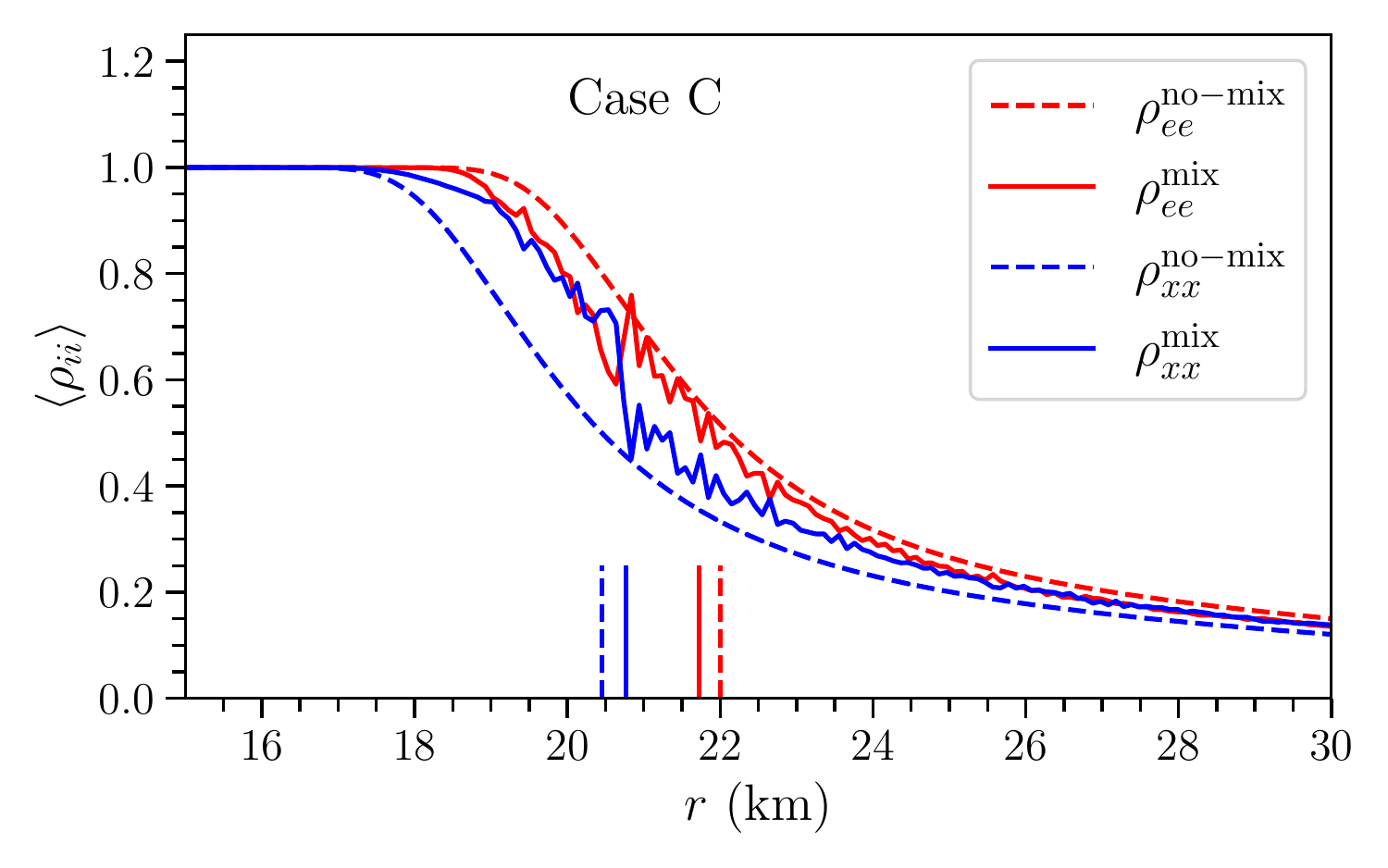}
\caption{Radial profile of the steady state angle-averaged  $\rho_{ee}$ (in red) and $\rho_{xx}$ (in blue)  in the presence of advection and collisions for Cases A, B, and C from top to bottom respectively. 
The dashed lines show the angle averaged number densities in the absence of neutrino flavor transformation, while the solid lines show the same with neutrino flavor transformation. The vertical lines mark the smallest radius at which the condition in  Eq.~\ref{ff} is fulfilled. While flavor equipartition is achieved in Case A, flavor equipartition is not reached in Cases B and C. In all cases, the neutrino decoupling surfaces are affected by flavor conversion according to the strength of the collision term.}
\label{angint}
\end{figure}

The spread of flavor transformed neutrinos from one radius to another is visible in the angle averaged neutrino occupation number as a function of the radius. We define the  (quasi) steady state angle averaged neutrino number density as follows: 
\begin{eqnarray}
\langle \rho_{ii}(r) \rangle = \frac{\int \rho_{ii}(r,\cos\theta) d\cos\theta}{\int d\cos\theta}\ . 
\end{eqnarray}
Figure~\ref{angint} shows the radial profiles of  $\langle \rho_{ii}(r) \rangle$ for Cases A, B, and C. Note that in the absence of advection and collisions,  unitary evolution dictates that  $\langle \rho_{ee}^{\textrm{mix}} \rangle$ lies in between  $\langle \rho_{ee}^{\textrm{no-mix}} \rangle$ and $\langle \rho_{xx}^{\textrm{no-mix}} \rangle$ 
for all radii. However, due to the presence of advection and collision, Fig.~\ref{angint} shows that this no longer holds; see for example Case B at $r \simeq 18$~km. 

It is worth stressing that it is not possible to disentangle the effects of collisions and advection in our formalism. 
Nevertheless,  their interplay ensures that neutrinos see different angular distributions over the simulation time, thus capturing the essence of the effects highlighted in Refs.~\cite{Shalgar:2020wcx,Padilla-Gay:2020uxa}.

\subsection{Effects of flavor transformation on neutrino decoupling}
As discussed in Ref.~\cite{Shalgar:2022rjj}, the angle averaged number density of neutrinos of various flavors offers insight into the effect of flavor transformation on neutrino decoupling.
In order to predict the region where decoupling approximately occurs, we consider the radius at which the flux factor for $\nu_i$ at the time when the quasi steady state configuration is reached,
\begin{eqnarray}
\mathcal{F}_{\nu_{i}}(r) = \frac{\int_{-1}^{1} \rho_{ii}(r, \cos\theta) \cos\theta d\cos\theta}{\int_{-1}^{1} \rho_{ii}(r, \cos\theta) d\cos\theta} \simeq \frac{1}{3}\ .
\label{ff} 
\end{eqnarray}
We  use this as an indicator of the effective decoupling radius since, 
in the coupled region, the neutrino angular distribution is isotropic, and the numerator vanishes. 
In the completely decoupled region, neutrinos are forward peaked and $\cos\theta \approx 1$ for all neutrinos; hence the flux factor is thus equal to unity. 

Figure~\ref{angint} shows the region where $\mathcal{F}_{\nu_{i}}(r,t) =1/3$ for  Cases A, B, and C, generalizing the findings of Ref.~\cite{Shalgar:2022rjj}. However, as visible in some panels of Fig.~\ref{angint}, changes in the angle averaged neutrino occupation number are not always directly correlated to changes in the decoupling radius. This is a consequence of the non-trivial angular distributions arising from neutrino flavor transformation.

\subsection{Interplay among flavor transformation, collisions, and advection}
The presence of collisions and advection redistributes neutrinos over the angle bins, as also found in Refs.~\cite{Shalgar:2020wcx,Shalgar:2019qwg}. 
In the absence of collisions and advection,  flavor transformation is predominantly present in a narrow angular range around the ELN crossing. 
Moreover, in the regions where neutrino flavor transformation is present, the angular structure becomes progressively finer with time. Because of this, more angle bins are required as the simulation time increases, when $\omega \neq 0$~\cite{Shalgar:2020xns,Johns:2019izj, Johns:2020qsk}. However, the growth of structure at smaller and smaller scales is suppressed due to the presence of collisions and advection (see also Appendix~\ref{app:num} and Ref.~\cite{Shalgar:2022rjj}). This is not surprising 
and it is due to the fact that the collision term redistributes neutrinos across angles. More importantly, in the present case, neutrinos at different radii undergo flavor transformation at different angles at a given time. The advective term mixes the angular distribution at various radii as neutrinos travel, reducing the number of angular bins required to reach angular convergence. 

The presence of flavor transformation in the angular range where $\nu_{e}$ and $\bar{\nu}_{e}$ are approximately equal is a consequence of the assumption of azimuthal symmetry. In the absence of azimuthal symmetry, the flavor transformation is not necessarily correlated to the angular region in the proximity of the ELN crossing~\cite{Shalgar:2021oko}. 

Recent literature has speculated that flavor equipartition or depolarization may be a general outcome of fast flavor evolution~\cite{Wu:2021uvt, Richers:2021nbx,Bhattacharyya:2020jpj}. 
Our findings suggest that this is not the case in our setup. Although we find equipartition in Case A, as shown in the top panel of Fig.~\ref{angint}, this is not true for Cases B and C. For Case A, flavor equipartition is reached because the occupation numbers of $\nu_{e}$ and $\bar{\nu}_{e}$ are very similar in the classical steady state configuration. It is also important to note that this finding in our simulation setup is also linked to the fact that 
   the lepton number in the neutrino sector is not conserved in our simulations due to absorption and emission collisional terms.

\section{Conclusions}
\label{sec:outlook}

Understanding the evolution of neutrino flavor in dense media remains an active subject of research. In this work, expanding on Ref.~\cite{Shalgar:2022rjj}, we investigate the flavor evolution for three different parametrizations of the collision term, consistently treating collision, advection, and flavor transformation. We rely on a spherically symmetric simulation shell and assume that all neutrinos have the same energy for simplicity. We populate the simulation shell through collisions and in the absence of flavor conversion, a steady state configuration is reached. In the presence of flavor transformation, flavor mixing spreads across angular and radial regions because of the dynamical effects induced by advection and collisions until a quasi steady state configuration is reached.

While in the literature flavor equipartition or depolarization is often presented as a  general outcome of fast flavor conversion, we find that  flavor equipartition is not achieved in general. Moreover, in the literature, it has been classically considered that fast flavor transformations could occur in the region of high density of neutrinos, while slow flavor collective transformations occur at larger radii and smaller densities; however, we find that an overlap between slow and fast flavor conversion could take place. This gives rise to a completely new phenomenology of neutrino self-interactions, yet to be explored. 

Our work highlights the dynamical interplay among flavor conversion, advection, and collisions. In particular, we find that flavor conversion spreads across angular modes and in a larger spatial range, instead of remaining clustered in the proximity of the ELN crossing. On the other hand, the dynamical interplay among flavor conversion, advection, and collisions hinders the cascade of flavor structure to small scales, otherwise expected~\cite{Shalgar:2020xns,Johns:2019izj, Johns:2020qsk}, and smears the quasi steady state distributions.

Because of the numerical challenges, our model includes some simplifications. The ones which further need to be relaxed concern the dependence on energy of flavor transformation and  the collision term. 
In fact, fast flavor conversion is by itself not sensitive to neutrino energy, but slow flavor conversion is. Hence, a non-trivial interplay between collisions and slow flavor conversion may exist. 

This work highlights the fascinating nature of neutrino self-interaction in compact astrophysical sources and the non-trivial interplay of the flavor conversion physics with neutrino advection and collisions. As such, our findings give a glimpse of flavor phenomenology that could have potentially interesting implications for the physics of compact sources and remains to be explored.

\acknowledgments
We would like to thank Rasmus S.L.~Hansen and Christopher Rackauckas for insightful discussions.
We acknowledge support from the Villum Foundation (Project No.~13164), the Danmarks Frie Forskningsfonds (Project No.~8049-00038B),  the MERAC Foundation, and the Deutsche Forschungsgemeinschaft through Sonderforschungbereich
SFB~1258 ``Neutrinos and Dark Matter in Astro- and
Particle Physics'' (NDM).


\appendix

\section{Numerical convergence}\label{app:num}
\begin{figure}[b]
\includegraphics[width=0.49\textwidth]{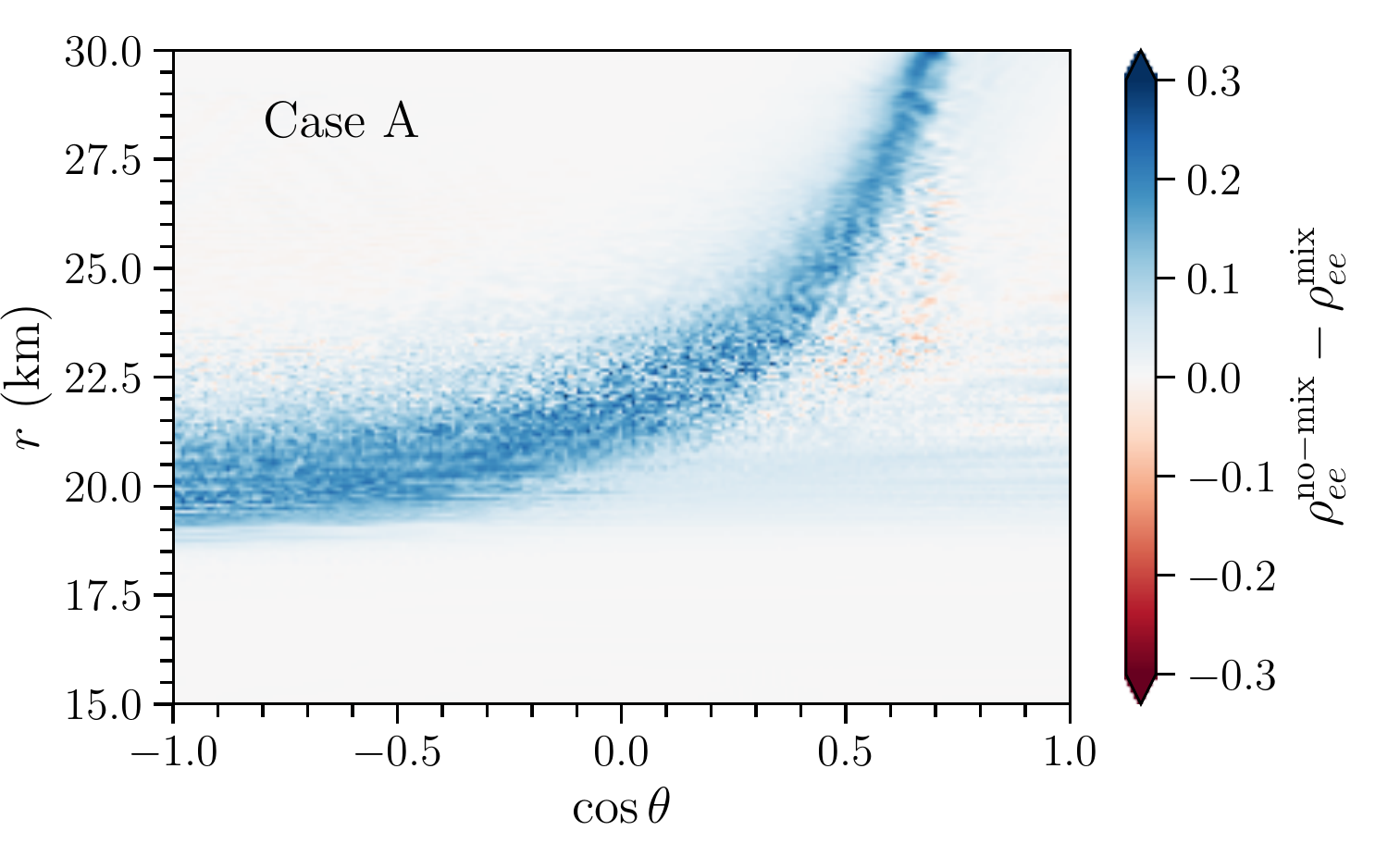}
\includegraphics[width=0.49\textwidth]{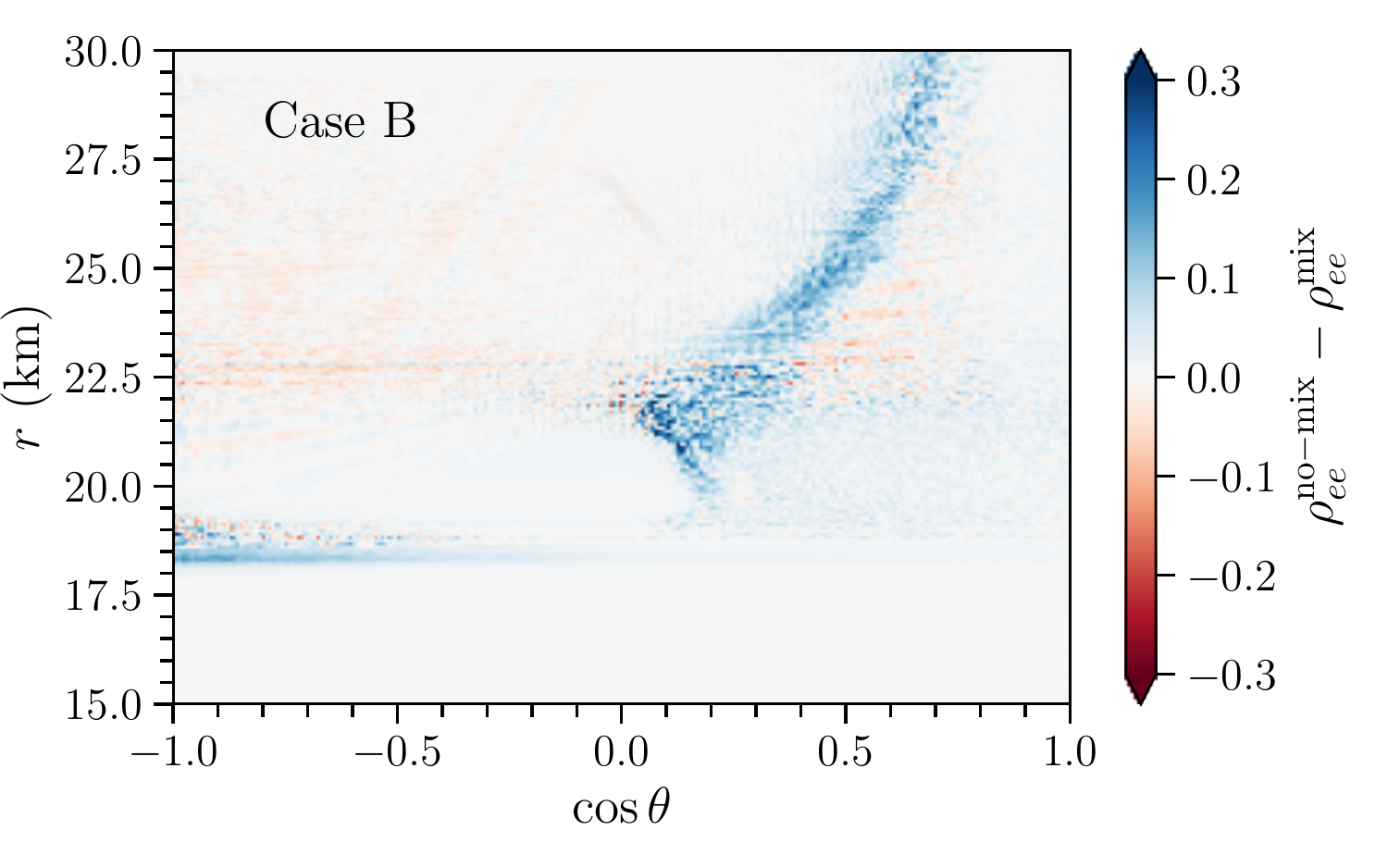}
\includegraphics[width=0.49\textwidth]{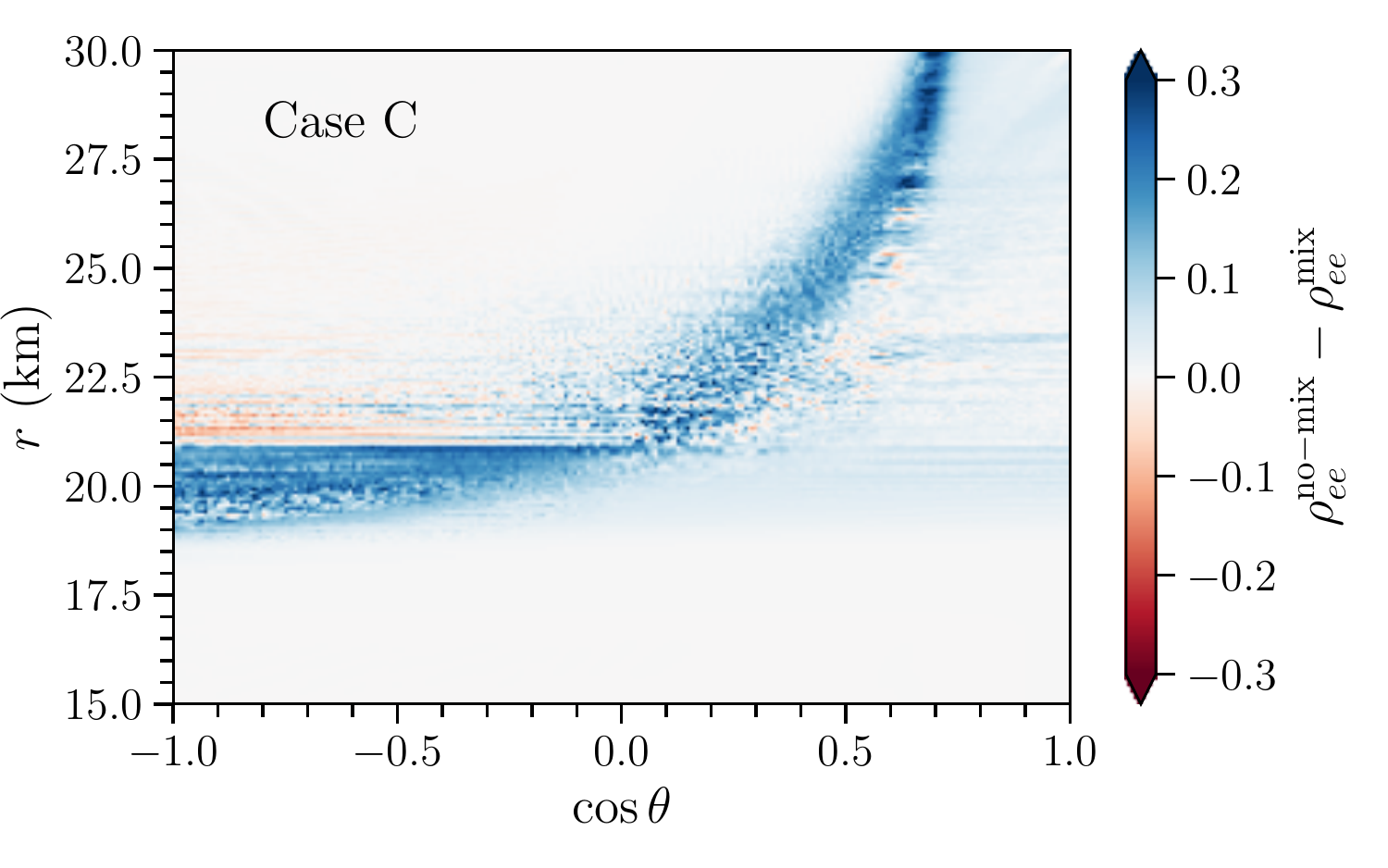}
\caption{Same Fig.~\ref{collandadvec},  but with $200 \times 200$ bins. The results are in agreement with the ones in Fig.~\ref{collandadvec}.}
\label{Fig2app}
\end{figure}

To prove numerical convergence, Fig.~\ref{Fig2app} shows the  analogous of Fig.~\ref{collandadvec}, but with simulations obtained by using  $200 \times 200$ bins, while keeping all other inputs unchanged. One can see that the agreement between the two figures is excellent. 

\begin{figure*}
\includegraphics[width=0.49\textwidth]{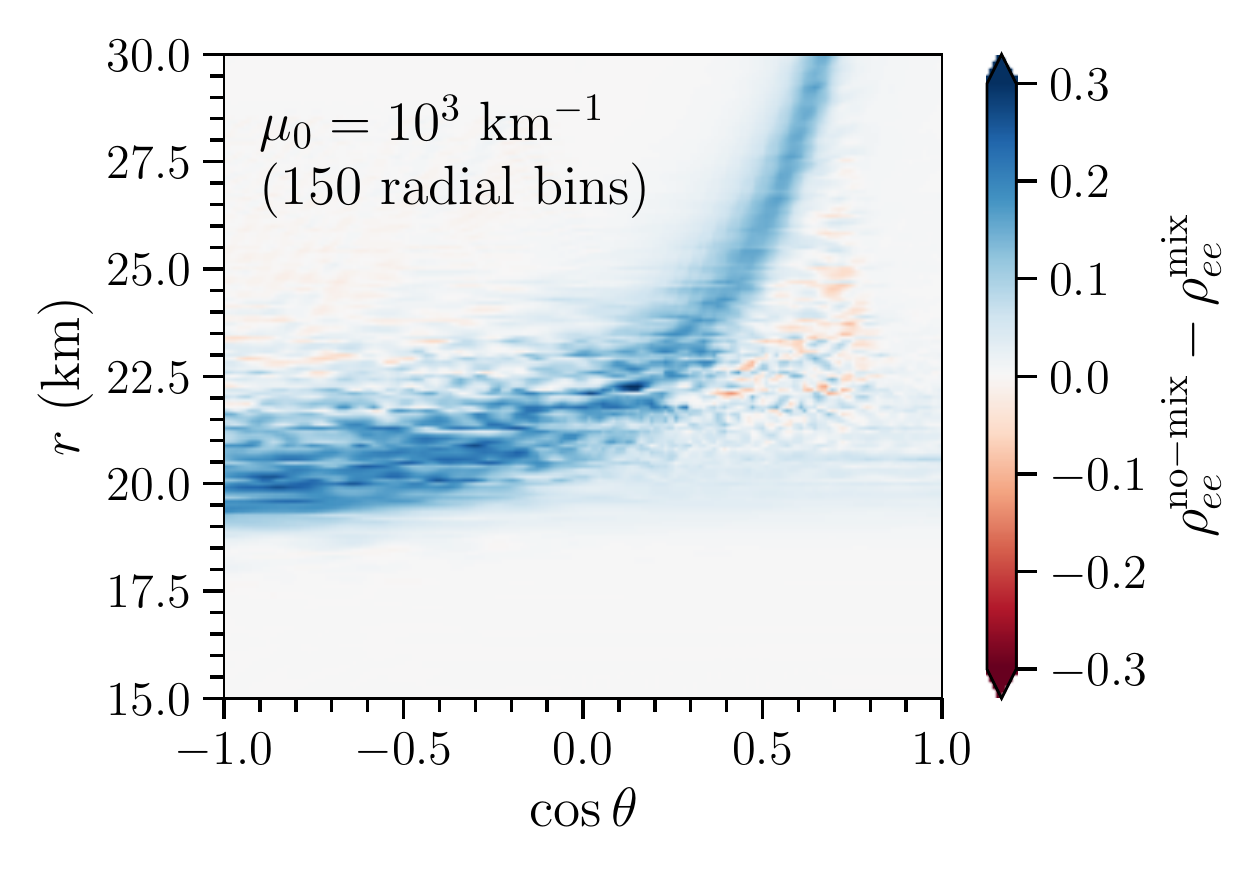}
\includegraphics[width=0.49\textwidth]{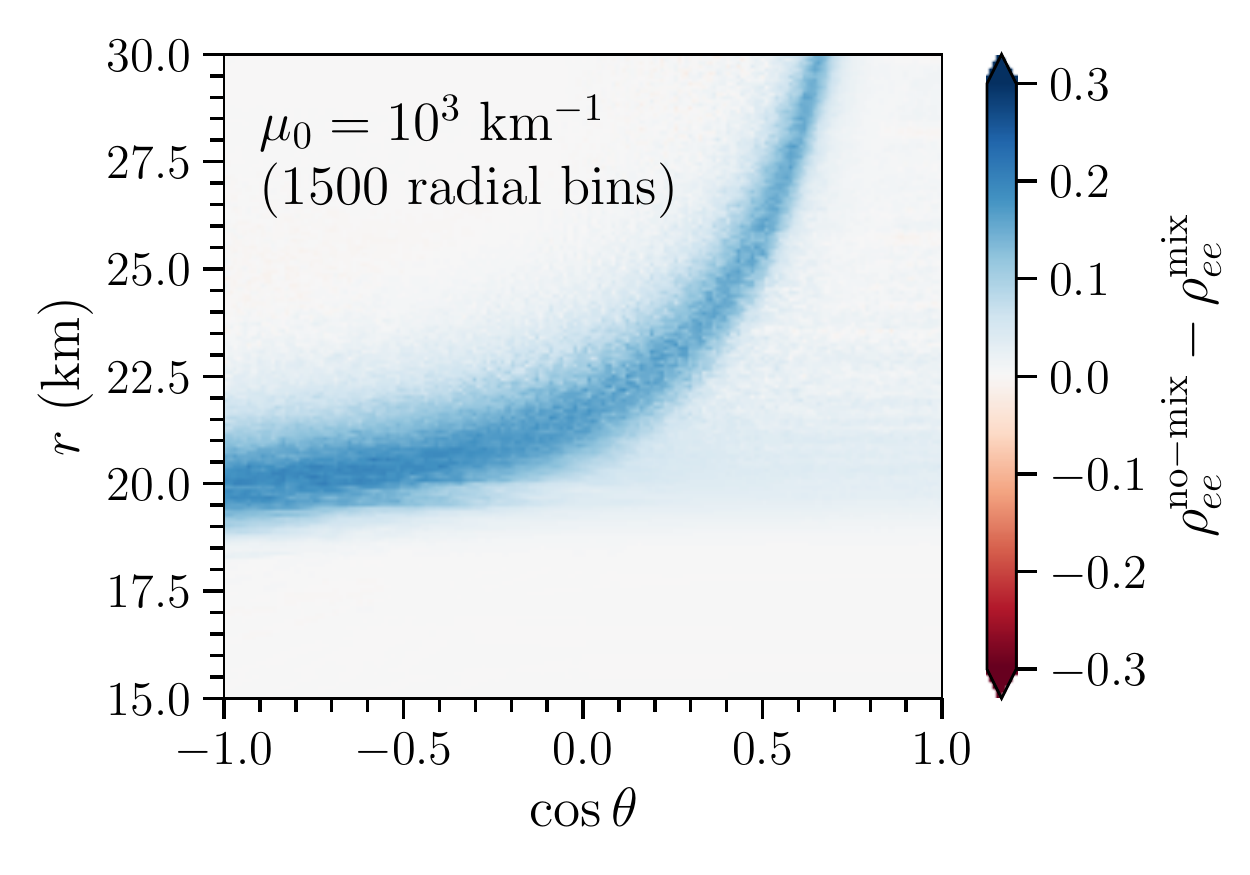}\\
\includegraphics[width=0.49\textwidth]{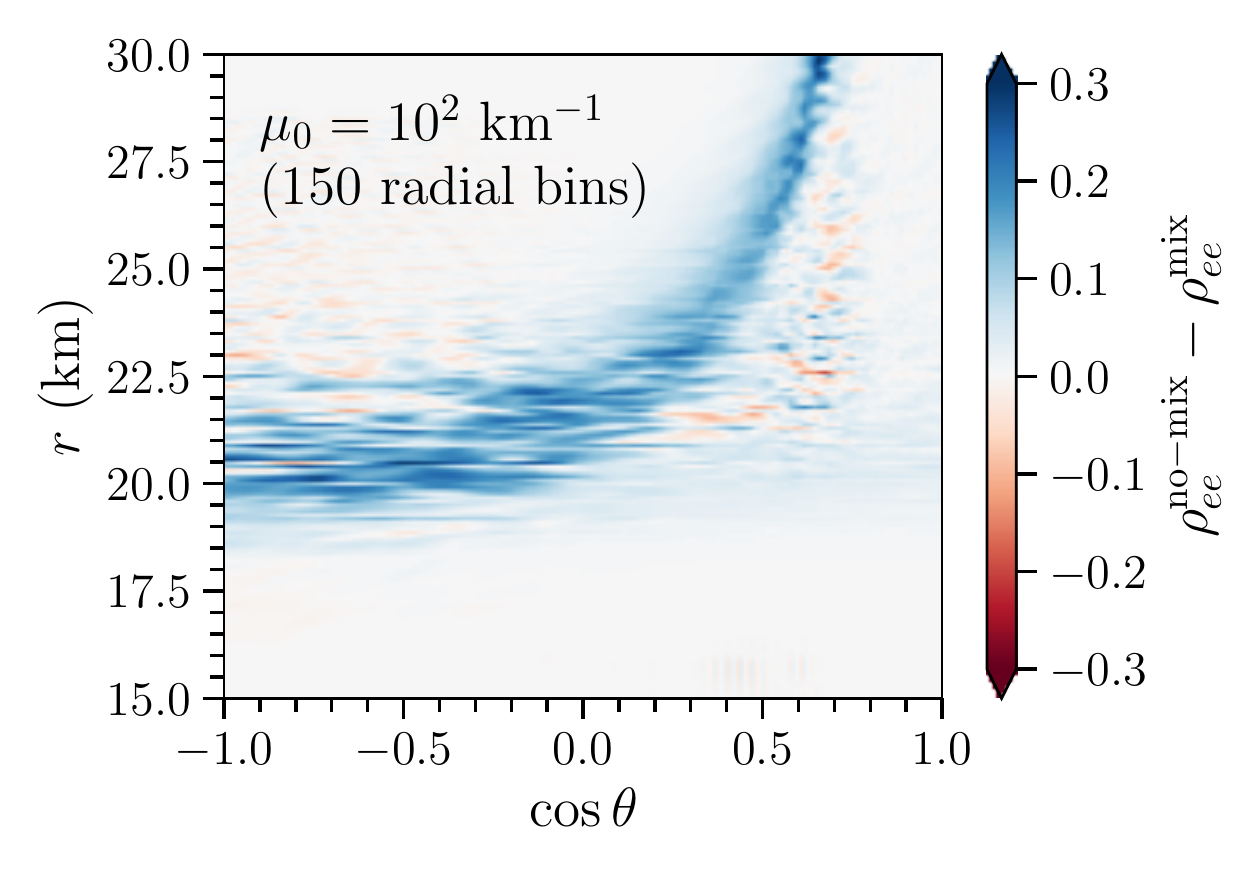}
\includegraphics[width=0.49\textwidth]{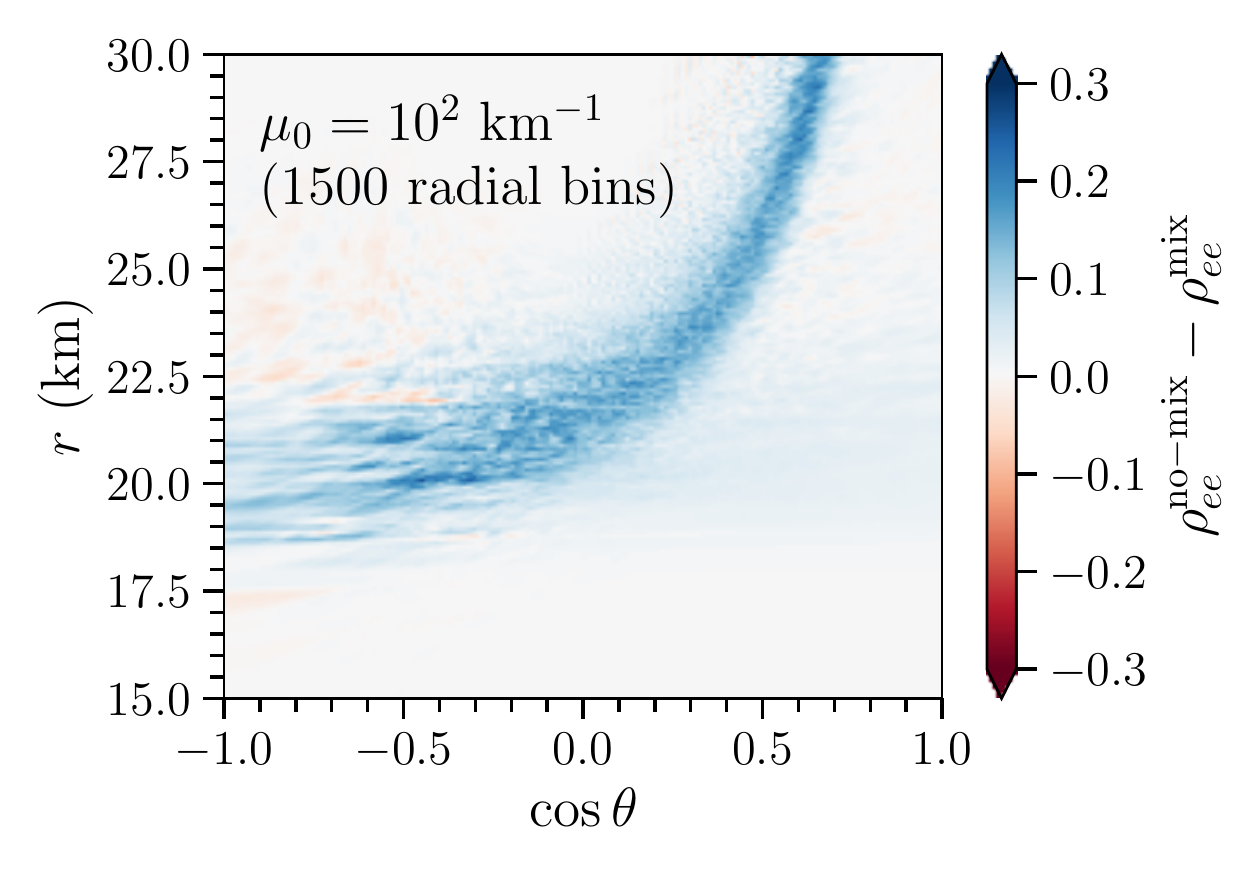}
\caption{Same as Fig.~\ref{collandadvec}, but  for Case A and $\mu_{0}=1000$ km$^{-1}$ (top panels) as well as $\mu_{0}=100$ km$^{-1}$ (bottom panels),  with 150 (left) and 1500 (right) radial bins. The average relative error  between the simulations with different number of radial bins  is $1.16\%$ for $\mu_{0}=1000$ km$^{-1}$ and $2.25\%$ for $\mu_{0}=100$ km$^{-1}$. 
}
\label{Fig1app}
\end{figure*}

Naively, one might expect that the simulation radial resolution should be dictated by $1/\mu_0$ (as it would be the case in the absence of advection, collisions, and the vacuum term). However this is not the case in our simulation setup, because of the inclusion of neutrino advection and collisions. To this purpose, we consider two simulation sets, one with $\mu_0=1000$ km$^{-1}$ and the other one with $\mu_0=100$ km$^{-1}$, and for each  $\mu_0$, we run two simulations, one  with  $150$  and one with $1500$ radial bins for Case A, while all other inputs are kept unchanged. 
Figure~\ref{Fig1app} shows the results: while minor differences are appreciable, the overall trend is robust. The cascade of the large scale inhomogeneous modes to small scales occurs  at  a rate that advection and collisions are not able to erase  only for  $\mu_0=1000$ km$^{-1}$.  For  $\mu_0=100$ km$^{-1}$ and 1500 radial bins, we resolve resolve spatial scales of order $\mu_0^{-1}$ and show that our conclusions are not affected. 

\begin{figure}
\includegraphics[width=0.49\textwidth]{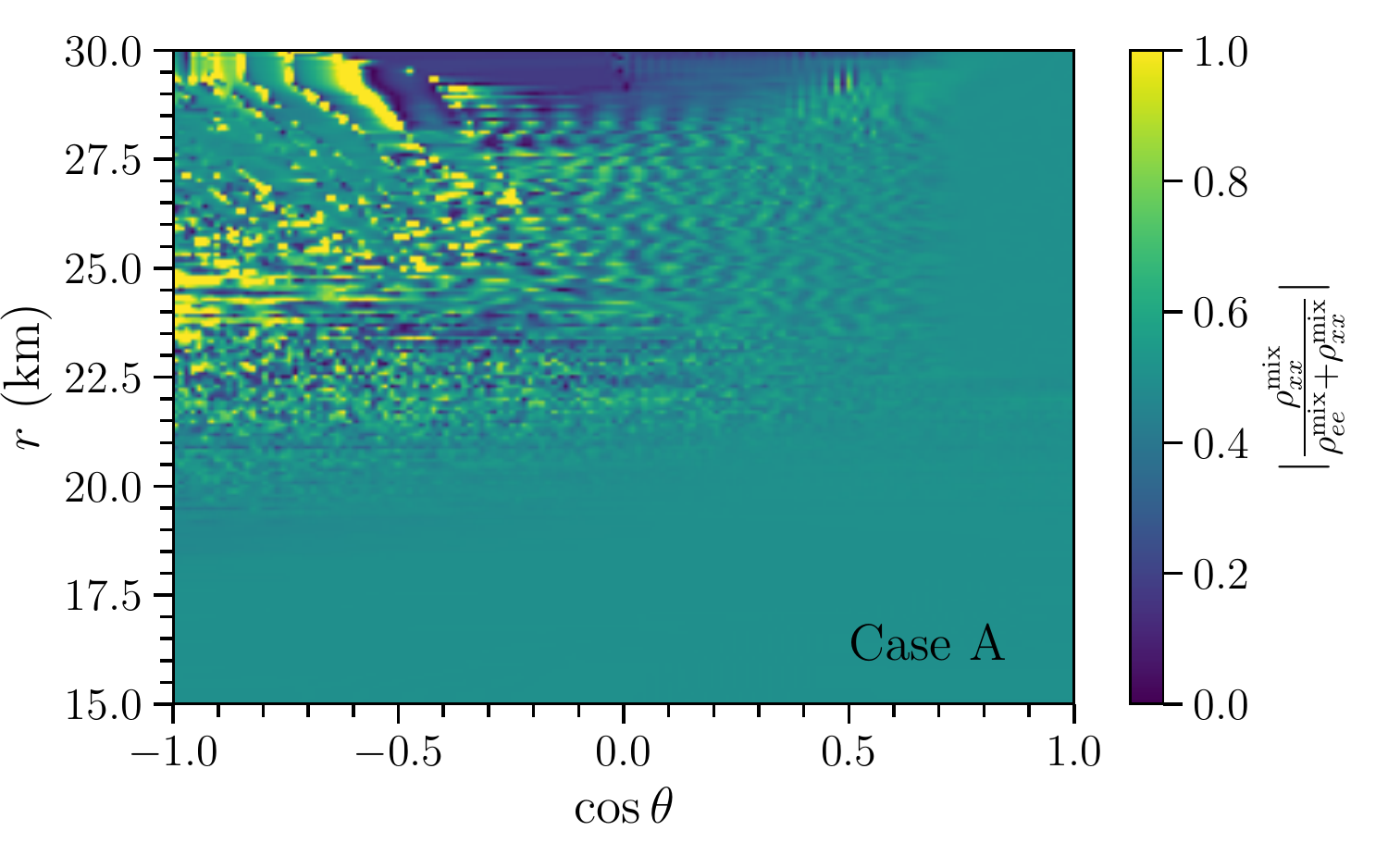}
\includegraphics[width=0.49\textwidth]{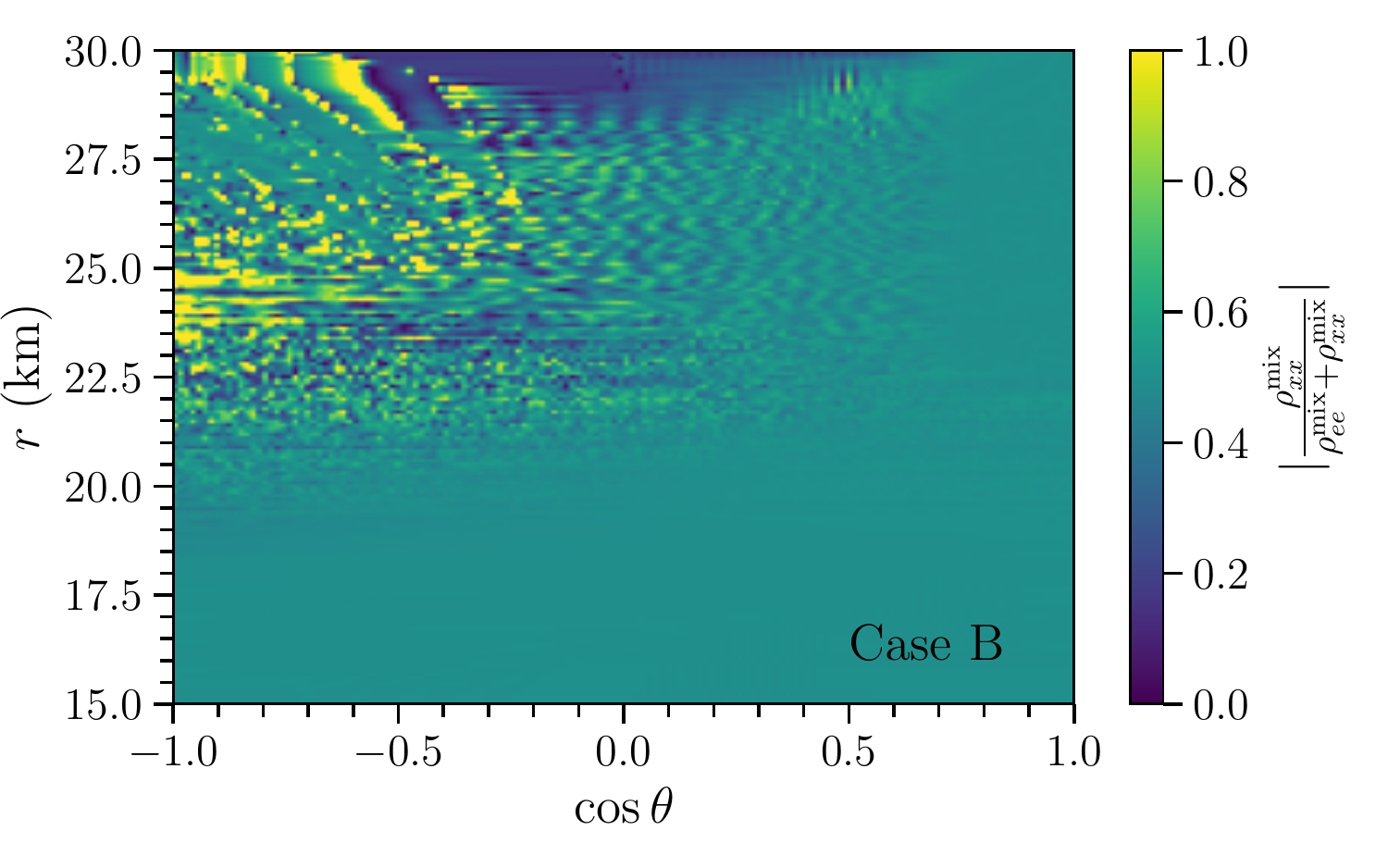}
\includegraphics[width=0.49\textwidth]{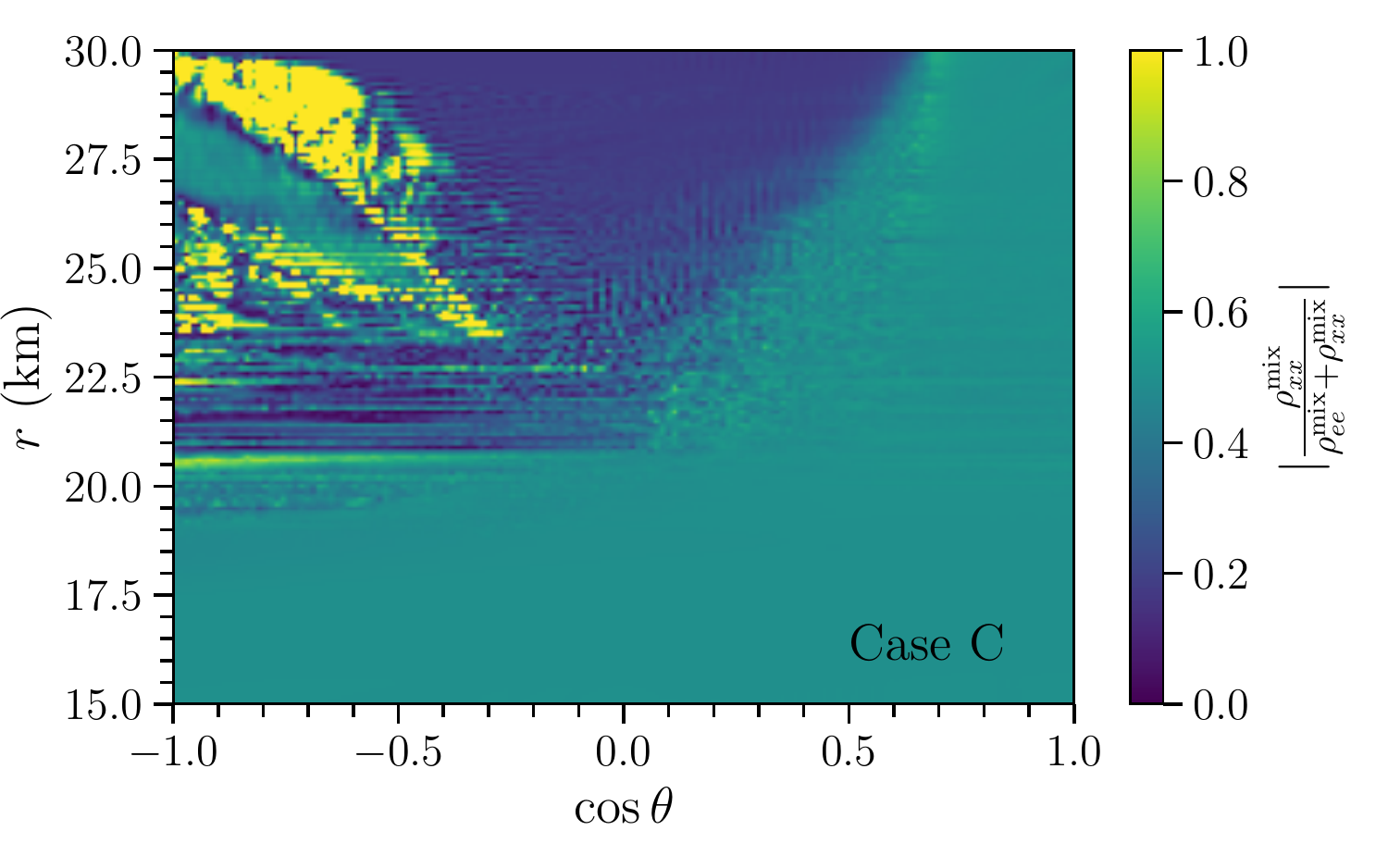}
\caption{Heatmap of $\left| {\rho^{\rm mix}_{xx}}/{(\rho^{\rm mix}_{ee}+\rho^{\rm mix}_{xx})} \right|$ after flavor conversion in the plane spanned by $\cos\theta$ and $r$ for Cases A, B, and C for the default value of $\mu_{0}=10^{4}$ km$^{-1}$.}
\label{FigSp}
\end{figure}

\begin{figure*}
\includegraphics[width=0.49\textwidth]{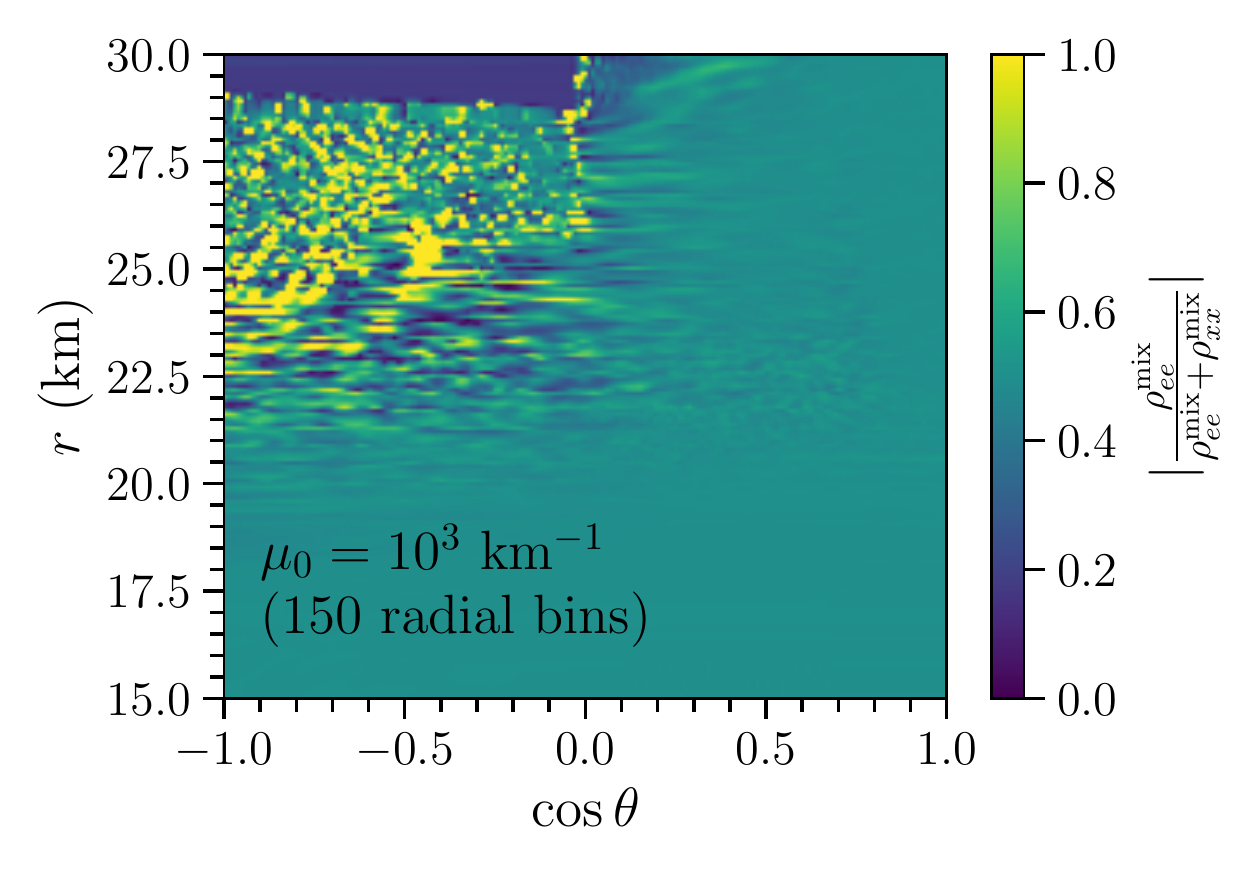}
\includegraphics[width=0.49\textwidth]{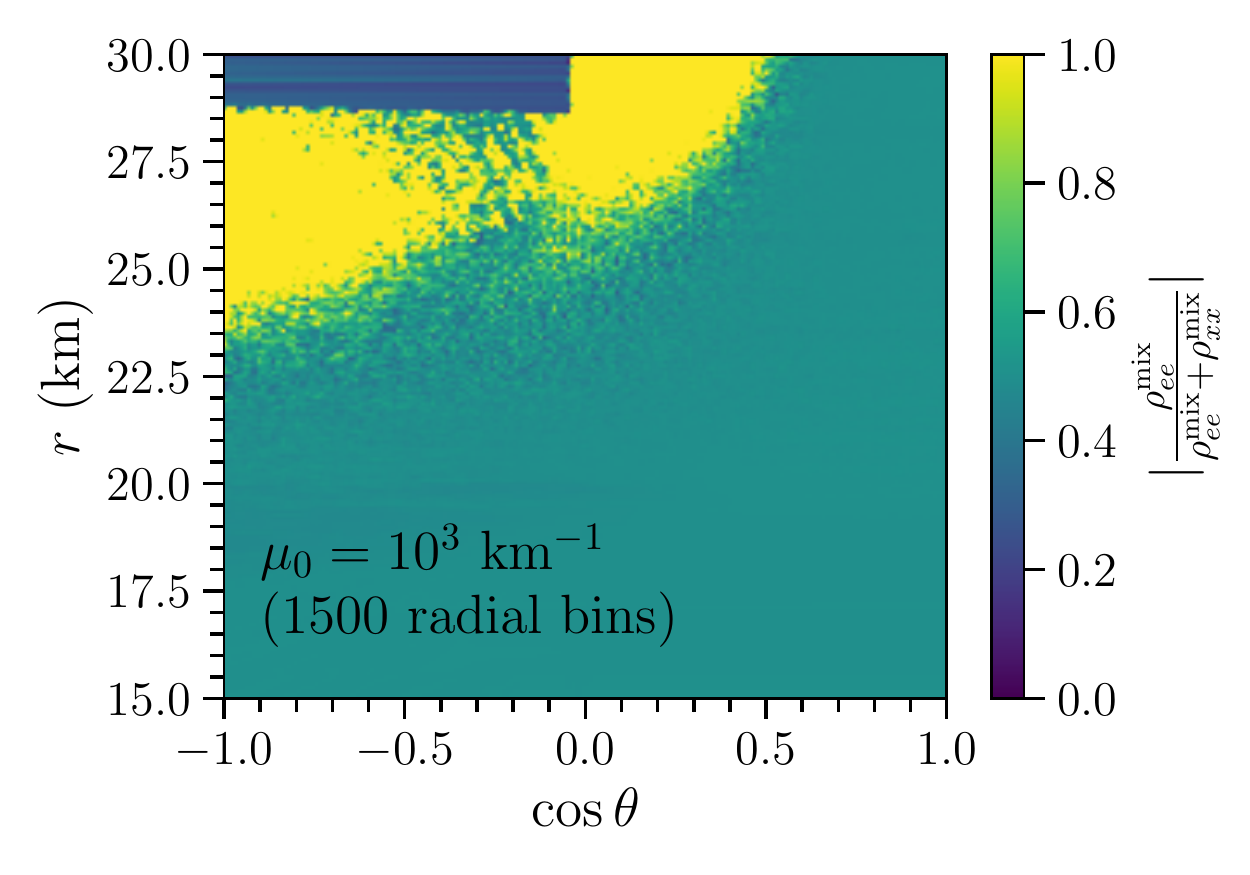}
\includegraphics[width=0.49\textwidth]{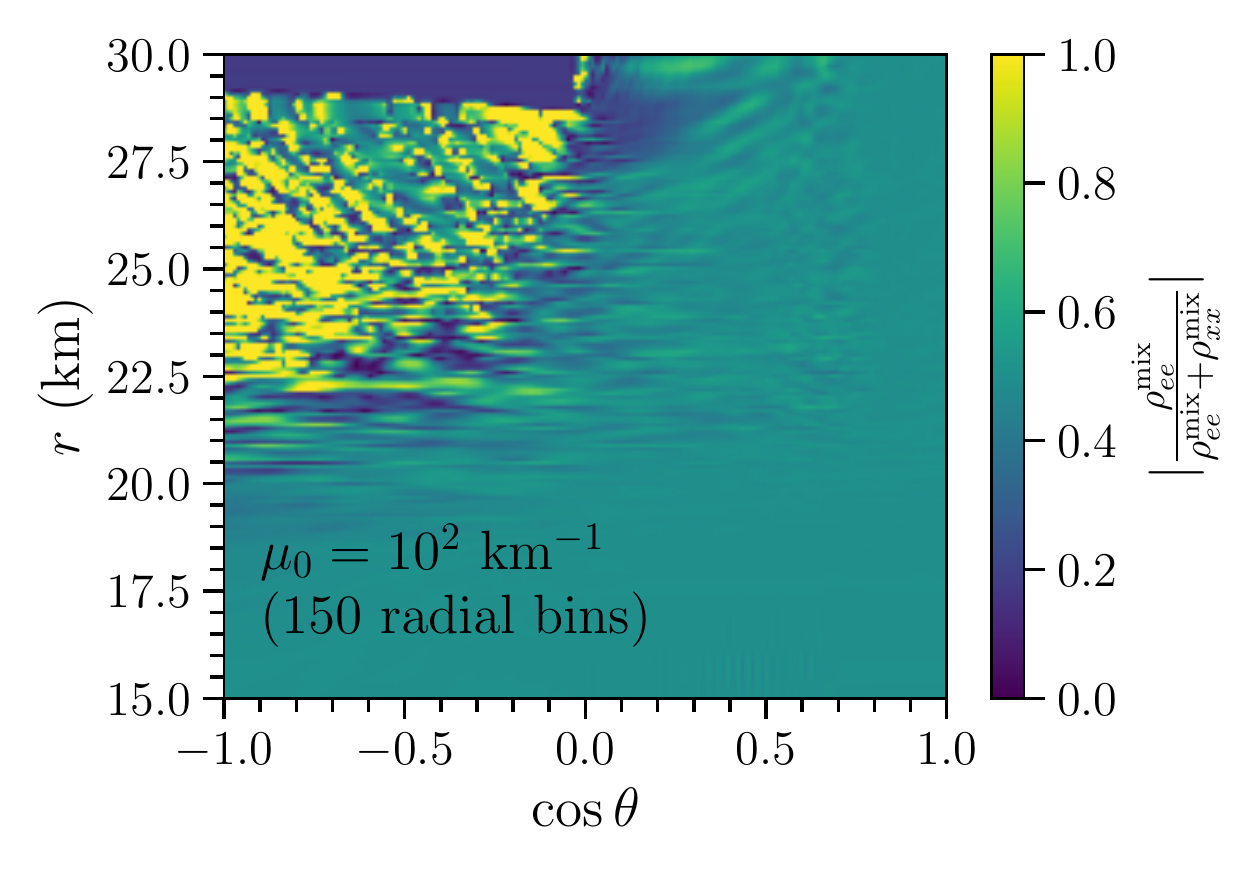}
\includegraphics[width=0.49\textwidth]{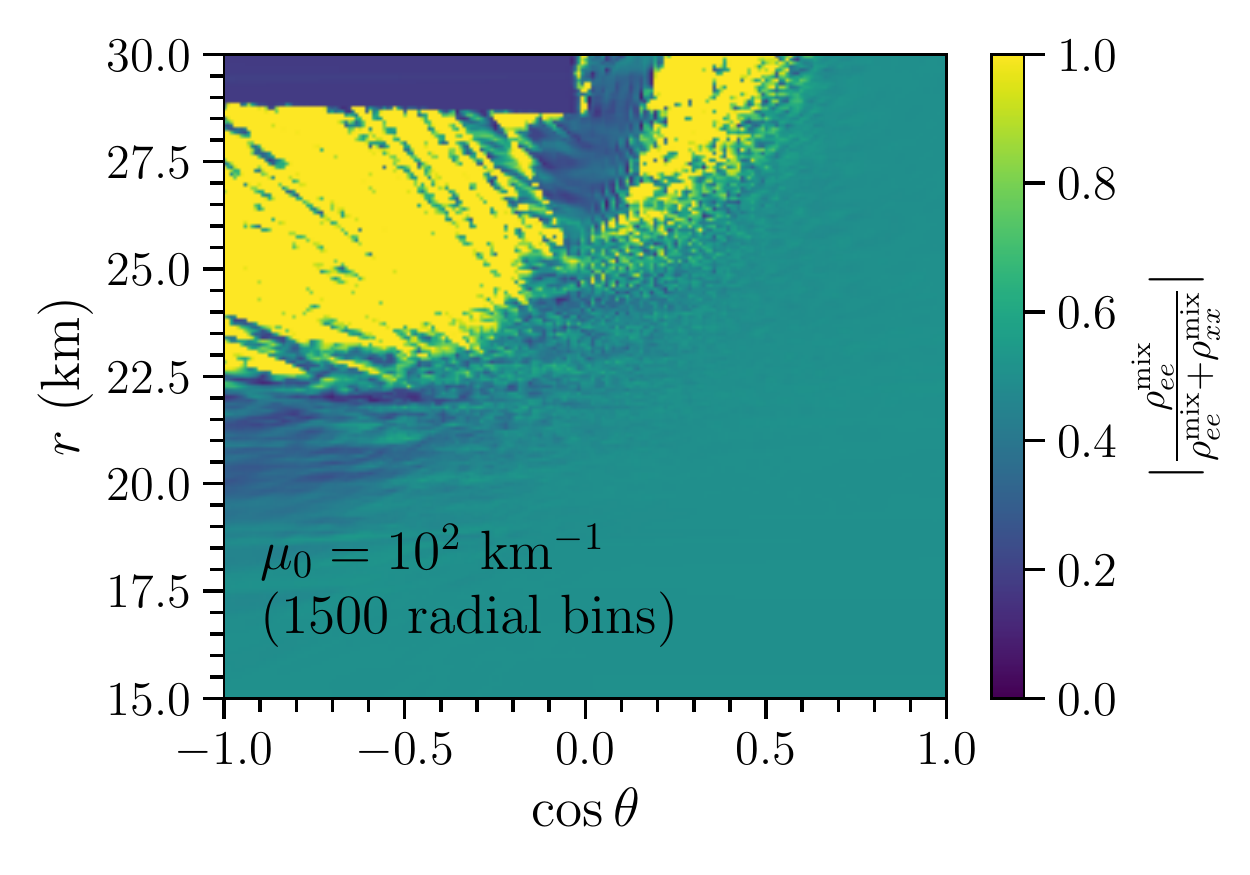}
\caption{Heatmap of $\left| {\rho^{\rm mix}_{xx}}/{(\rho^{\rm mix}_{ee}+\rho^{\rm mix}_{xx})} \right|$ after flavor conversion in the plane spanned by $\cos\theta$ and $r$ for Cases A for $\mu_{0}=10^{3}$ km$^{-1}$ (top panels) and $\mu_{0}=10^{2}$ km$^{-1}$ (bottom panels). The panels on the left have been obtained using 150 radial bins and the right panels have been obtained using 1500 radial bins. Note that the blue band in the top-left region is a numerical artifact, where the denominator is very small.}
\label{ratios}
\end{figure*}

To quantify the differences between the cases with different resolution in Fig.~\ref{Fig1app}, we compute the relative error  by coarse-graining the results with $1500$ radial bins and  averaging over batches of $10$ radial bins to compare the results with the ones from the numerical simulation with $150$ radial bins. At each radius, the error  is defined as relative error between the angle averaged population densities. We further average the relative error over all the radial bins to calculate the average relative error (see Ref.~\cite{Shalgar:2022rjj} for additional details). The average relative error at $t= 5 \times 10^{-5}$~s between the two simulations with different number of radial bins  is $1.16\%$ for $\mu_{0}=1000$ km$^{-1}$ and $2.25\%$ for $\mu_{0}=100$ km$^{-1}$. The figures demonstrate that the eventual existence of structures at small length scales of size $\mu_0^{-1}$  does not affect the results qualitatively even if the spatial resolution is not of that order. These findings are also in agreement with the ones presented in Refs.~\cite{Shalgar:2019qwg,Padilla-Gay:2020uxa,Shalgar:2020wcx}. Figure~\ref{FigSp} shows $\left| {\rho^{\rm mix}_{xx}}/{(\rho^{\rm mix}_{ee}+\rho^{\rm mix}_{xx})} \right|$ for Cases A, B, and C using the default value of $\mu_{0}=10^{4}$ km$^{-1}$ for the sake of completeness. In addition, Fig.~\ref{ratios} shows the same quantity for lower value of $\mu_{0}$ for Case A (see also Fig.~\ref{Fig1app}). Note that some of these plots should be interpreted with caution because of  numerical artifacts in the top-left region where the denominator is very small.

\bibliography{fullsolution.bib}
\end{document}